\documentclass[times,review,preprint,authoryear]{elsarticle}

\usepackage{jasr}
\usepackage{framed,multirow}

\usepackage{amssymb}
\usepackage{latexsym}

\usepackage{amsmath}
\usepackage{amsfonts}
\usepackage{amsthm}
\usepackage{newlfont}
\usepackage{mathrsfs}
\usepackage{float}
\usepackage{ragged2e}
\usepackage{xcolor}
\usepackage{longtable}
\usepackage{multirow}
\usepackage{colortbl}
\usepackage{makecell}
\usepackage{caption}
\usepackage{subcaption}

\usepackage{url}
\definecolor{newcolor}{rgb}{.8,.349,.1}

\usepackage[citebordercolor=white]{hyperref}

\journal{Advances in Space Research}

\newcommand{\ra}{\ddot{\mathbf{r}}}
\newcommand{\rv}{\dot{\mathbf{r}}}
\newcommand{\rhat}{\hat{\mathbf{r}}}
\newcommand{\nhat}{\hat{\mathbf{n}}}
\newcommand{\thetahat}{\hat{\boldsymbol{\theta}}}
\newcommand{\phihat}{\hat{\boldsymbol{\varphi}}}

\newcommand{\J}{\mathbf{J}}
\newcommand{\thetadot}{\dot{\theta}}

\newcommand{\phidot}{\dot{\varphi}}
\newcommand{\rdot}{\dot{r}}

\newcommand{\h}{\mathbf{h}}
\renewcommand{\r}{\mathbf{r}}

\newcommand{\tbeta}{\tilde{\beta}}

\renewcommand{\d}{\mathrm{d}}

\newcommand*{\Scale}[2][4]{\scalebox{#1}{$#2$}}%

\begin{document}

\verso{Jeric Garrido \& Jose Perico Esguerra}

\begin{frontmatter}

\title{A surface constraint approach for solar sail orbits}%

\author[1]{Jeric \snm{Garrido}\corref{cor1}}
\cortext[cor1]{Corresponding author} 
\ead{jgarrido@nip.upd.edu.ph}
\author[1]{Jose Perico  \snm{Esguerra}\fnref{fn1}}
\ead{jhesguerra1@up.edu.ph}

\address[1]{National Institute of Physics, University of the Philippines, Diliman, Queezon City 1101, Philippines}

\received{}
\finalform{}
\accepted{}
\availableonline{}
\communicated{}

\begin{abstract}
In this paper, a surface geometric constraint approach is used in designing the orbits of a solar sail. We solve the solar sail equation of motion by obtaining a generalized Laplace-Runge-Lenz (LRL) vector with the assumption that the cone angle is constant throughout the mission. A family of orbit equation solutions can then be specified by defining a constraint equation that relates the radial and polar velocities of the spacecraft and is dependent on the geometry of the surface where the spacecraft is expected to move. The proposed method is successfully applied in the design of orbits constrained on cylinders and to displaced non-Keplerian orbits.
\end{abstract}

\begin{keyword}
\KWD Orbit determination\sep Surface geometric constraint\sep Laplace-Runge-Lenz vector  	
\end{keyword}

\end{frontmatter}

\section{Introduction}
\label{sec:intro}

Solar sailing is a method of space propulsion where a spacecraft is propelled and sent potentially anywhere using the solar radiation pressure coming from the sun \citep{mcinnes2004solar,vulpetti2014solar,gong2019review}. The large $\Delta V$ changes in the solar sailing spacecraft caused by the continuous thrust generated by solar radiation pressure makes solar sails viable in missions beyond the solar system and involving deep space 
\citep{vulpetti2012fast,lingam2020propulsion}, missions monitoring a planet's activity \citep{lappas2009practical,macdonald2007geosail,ozimek2009design} or harnessing materials from asteroids \citep{morrow2001solar,peloni2016solar,song2019solar}. 

In solar sailing mission applications, it is necessary to obtain an orbit equation that characterizes the trajectory of the sail. To determine these families of orbit equations, one needs to integrate the equation of motion subjected to the initial and terminal conditions that are defined based on the intended mission. In some cases, a closed form expression for the orbit equation of the solar sail can be determined \citep{petropoulos2002review}. Examples of these closed-form solutions include the logarithmic spiral trajectory, displaced non-Keplerian orbits \citep{mcinnes1998dynamics,mckay2011survey}, and orbits constrained on cylinders and spheres \citep{heiligers2014new}.

For conservative systems, it suffices to determine their constants of motion such as angular momentum and energy to deduce their dynamics. However, for low thrust spacecraft systems like solar sails, using this approach poses a challenge since the equation of motion does not give the usual first integrals of motion \citep{conway2010spacecraft}. It is still possible though to use a generalized version of these conserved quantities for orbit determination. \citet{leach2003generalisations} have shown that for equations of motion where either the direction or the magnitude of the angular momentum is conserved, a generalized Laplace-Runge-Lenz (LRL) vector can be determined and integrated, yielding an orbit equation. This approach was used to calculate the orbit equations of the Kepler problem with drag and systems in full three dimensional motion \citep{gorringe1987conserved,leach1987first}. 

One way to design orbit equation solutions of continuously-propelled systems like solar sails is through the shape-based approach \citep{vasile2007optimality}. In the shape-based approach \citep{petropoulos2003shape,petropoulos2004shape}, the trajectory equation or shape of the spacecraft is assumed \textit{a priori}, and then the free parameters are adjusted to match with the equation of motion and boundary conditions. Different shapes like the exponential sinusoid \citep{petropoulos2003shape,petropoulos2004shape} and the inverse quintic polynomial \citep{wall2009shape,wall2008shape} have been used for orbit determination and/or optimization. The shape-based approach was successful in solving the Lambert's problem for exponential sinusoids \citep{izzo2006lambert} as well as solving three dimensional rendezvous trajectories \citep{vasile2007optimality}. A generalized shape based formulation was proposed by  \citet{taheri2012shape} which utilizes Fourier series and by \citet{fan2020initial} which uses Bezier curves to rapidly generate low thrust trajectories. 

The objective of this paper is to propose an alternative approach of orbit determination through the use of a surface geometric constraint. We obtain orbit equation solutions from the equation of a solar sail by modifying the method by Leach \textit{et al.}. Instead of assuming an ansatz of the orbit like in the shape-based approach, the proposed approach first gives us a set of possible families of orbits based on the equation of motion. We can then choose the orbit type from these families or orbits by specifying the surface geometric constraint-- a function related to the surface where the spacecraft is expected. 

The paper is organized as follows: Section \ref{sec:geom_cons} outlines the general approach of our orbit determination scheme.  We then apply the approach to two different cases: first on orbits constrained on cylinders (Section \ref{sec:applns_cyl}) and second on displaced non-Keplerian orbits (Section \ref{sec:displaced-NKO}). We characterize these orbits based on how the sail's attitude changes in space.

\section{Surface Constraint Approach}
\label{sec:geom_cons}

\begin{figure}[h!]
	\centering
	\includegraphics[scale=0.35]{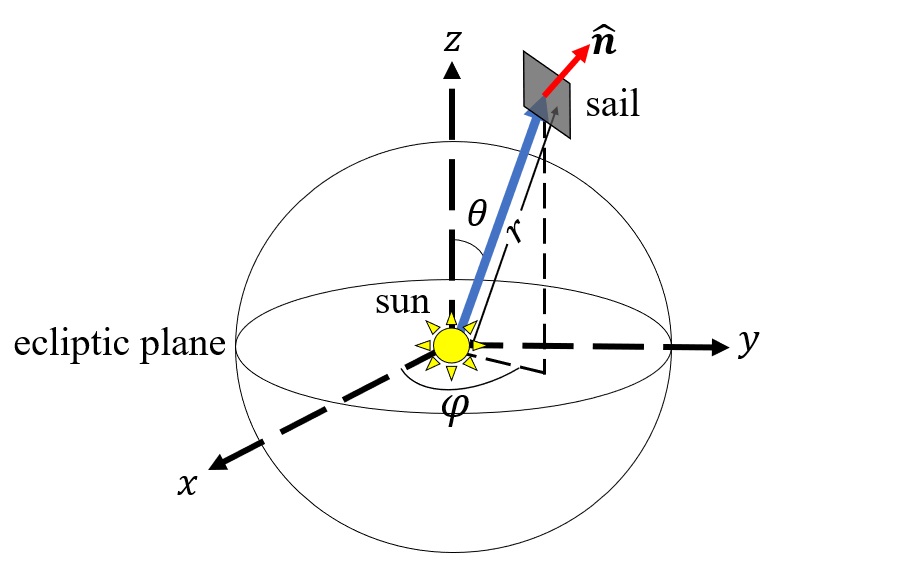}
	\caption{The coordinate system to be used in this paper. The vector $\nhat$ is the unit vector to the effective area of the sailing spacecraft.}
	\label{fig:coord_system}
\end{figure}

In this section, we will obtain solar sail orbit equations on surfaces. Consider a rigid, flat, perfectly reflecting solar sailing spacecraft that moves in a three-dimensional sun-centered inertial reference frame. We use the simplifying assumption that the sun is a point-like source of radiation and is sufficiently far from the sail so that the sun's rays are nearly parallel when incident to it. The position of the sail is defined by three coordinates: the radial coordinate $r$ which is the radial distance from the sun, the polar angle $\theta$ which is the angle relative to the north ecliptic pole, and the azimuthal angle $\varphi$ as the angle relative to the position of the First Point of Aries taken in the counterclockwise direction. Using this coordinate representation, any vector can be written in terms of the unit vectors $\{\rhat,\thetahat,\phihat\}$. See Figure \ref{fig:coord_system}. 

\begin{figure}[h!]
	\centering
	\includegraphics[scale=0.4]{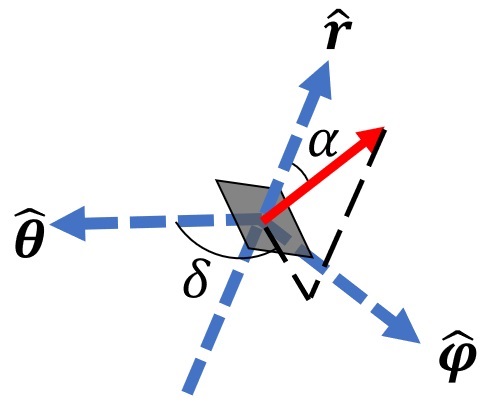}
	\caption{The cone angle $\alpha$ and the clock angle $\delta$ defined in the heliocentric spherical polar coordinate system.}
	\label{fig:cone-clock-angles}
\end{figure}

Let $\nhat$ be the unit normal vector to the sail's surface. Then, $\nhat$ can be represented in our spherical coordinate system by defining two angles, namely, the cone angle $\alpha$ as the angle between $\rhat$ and $\nhat$, and the clock angle $\delta$ which is the angle between $\thetahat$ and $\nhat$. From these definitions, we can express $\nhat$ in terms of the spherical unit vectors:
\begin{equation}
	\label{eq:nhat_def}
	\begin{split}
		\nhat&=\rhat n_{r}+\thetahat n_{\theta}+\phihat n_{\varphi}\\
		&= \rhat\cos\alpha+\thetahat\sin\alpha\cos\delta+\phihat\sin\alpha\sin\delta
	\end{split}
\end{equation}

The equation of motion of the sail is then given by
\begin{equation}
	\label{eq:EOM_3D}
	\ra=-\frac{\mu}{r^{2}}\rhat+\frac{\beta\mu}{r^{2}}\nhat(\nhat\cdot\rhat)^{2}
\end{equation}
where $\mu=GM_{\odot}$ is the solar gravitational parameter. We then solve equation \eqref{eq:EOM_3D} to obtain solutions of the form $r(\theta,\varphi)$.

\subsection{Generalized Laplace-Runge-Lenz vector}

An orbit equation can be determined from a generalized Laplace-Runge-Lenz (LRL) vector derived from the equation of motion \citep{leach2003generalisations}.  The LRL vector can be obtained by considering the antiderivative of $\rv\times\h$, where $\h$ is the specific angular momentum  of the sail. It is noteworthy that $\h$ is not conserved in this system since the time rate of change of the angular momentum, given by
\begin{equation}
	\dot{\h}=\frac{\beta\mu}{r}(\nhat\cdot\rhat)^{2}(\rhat\times\nhat),
\end{equation} 
is non-zero. 

To obtain a conserved vector, we cross multiply the equation of motion \eqref{eq:EOM_3D} with $\h$, that is
\begin{equation}
	\label{eq:eom_times_h}
	\left(\ra+\frac{\mu}{r^{2}}\rhat-\frac{\beta\mu}{r^{2}}\nhat(\nhat\cdot\rhat)^{2}\right)\times\h=0.
\end{equation}
Note that the first term in the above expression can be expanded using product rule. Hence,
\begin{equation}
	\rv\times\h=\frac{\d}{\d t}\left(\rv\times\h\right)-\rv\times\dot{\h}.
\end{equation}
Then, we express each of the resulting terms in equation \eqref{eq:eom_times_h} as a derivative of another function. One can easily verify that the second term of \eqref{eq:eom_times_h} reduces to
\begin{equation}
	\frac{\mu}{r^{2}}\rhat\times\h=-\frac{\d}{\d t}\left(\frac{\mu}{r}\r\right).
\end{equation}
Thus, equation \eqref{eq:eom_times_h} can be written as
\begin{equation}
	\frac{\d}{\d t}\left(\rv\times \h\right)-\frac{\d}{\d t}\left(\frac{\mu}{r}\r\right)-\rv\times\dot{\h}-\frac{\beta\mu}{r^{2}}\nhat(\nhat\cdot\rhat)^{2}\times\h=0
\end{equation}
To simplify the last two terms, we have to impose that \textit{the radial component of the unit vector, $n_{r}=\nhat\cdot\rhat$, is constant}. This is equivalent to fixing the cone angle throughout the mission. Because we have not imposed any condition on the polar and azimuthal components other than the unit normal vector being normalized, the sail may rotate relative to the sun line.

By imposing the constancy of $n_{r}$, one can show that
\begin{equation}
	\frac{\beta\mu}{r^{2}}(\nhat\cdot\rhat)^{2}(\nhat\times\h)=\frac{\d}{\d t}\left((\nhat\cdot\rhat)^{2}\nhat\right)
\end{equation}
and
\begin{equation}
	\rv\times\h=\beta\mu\frac{\d}{\d t}\left((\nhat\cdot\rhat)^{2}\nhat-(\nhat\cdot\rhat)^{3}\rhat\right)+\beta\mu\frac{\rdot}{r}(\nhat\cdot\rhat)^{2}\left(n_{\theta}\thetahat+n_{\varphi}\phihat\right).
\end{equation}
Hence, equation \eqref{eq:eom_times_h} becomes
\begin{equation}
	\label{eq:medyo_simp}
	\frac{\d}{\d t}\left(\rv\times \h-\mu\rhat+\beta\mu  n_{r}^{2}(2\nhat-n_{r}\rhat)\right)+\beta\mu n_{r}^{2}\frac{\rdot}{r}(n_{\theta}\thetahat+n_{\varphi}\phihat)=0.
\end{equation}
Taking the antiderivative of equation \eqref{eq:medyo_simp}, a conserved generalized Laplace-Runge-Lenz vector can be obtained:
\begin{equation}
	\label{eq:LRL_vector}
	\displaystyle{\J=\rv\times \h-\mu\rhat+\beta\mu n_{r}^{2}(2\nhat-n_{r}\rhat)+\beta\mu n_{r}^{2}\int\d t\frac{\rdot}{r}(n_{\theta}\thetahat+n_{\varphi}\phihat).}
\end{equation}
If the force exerted on the sail is purely gravitational, then this vector reduces to the Laplace-Runge-Lenz vector for Keplerian orbits. In principle, a dot product can be performed between the generalized LRL vector \eqref{eq:LRL_vector} and the position vector to obtain an orbit equation solution. However, the third term is difficult to evaluate for the general case. To avoid this difficulty, we will consider cases where this term simplifies due to the geometry of the problem.

\subsection{Surface Constraint Equation}
In some situations, we want our sail to be restricted on surfaces of revolution for which the radial position is dependent only on the polar angle. In these cases, we can define a constraint equation where we assume that the radial component of the sail's velocity is related to the polar component by some function $G$ that depends only on $\theta$. That is,
\begin{equation}
	\label{eq:constraint}
	\rdot=rG(\theta)\thetadot.
\end{equation}
Solving this equation, we see that
\begin{equation}
	\label{eq:orb_eqn_first_part}
	r(\theta)=r_{0}\exp\left(\int_{\theta_{0}}^{\theta}G(\nu)\d\nu\right).
\end{equation}
Observe that the radial orbit equation is independent of the azimuthal angle. 

We can also substitute the constraint equation \eqref{eq:constraint} to the LRL vector. If we assume that $n_{\theta}$ and $n_{\varphi}$ are functions of the polar angle, then the LRL vector can be written as
\begin{equation}
	\displaystyle{\J=\rv\times\h-\mu\rhat+\beta\mu n_{r}^{2}(2\nhat-n_{r}\rhat)+\beta\mu n_{r}^{2}\int_{\theta_{0}}^{\theta}\d\eta\ G(\eta)(n_{\theta}\thetahat+n_{\varphi}\phihat)}.
\end{equation}
We can take the unit vectors out of the integral by expressing them in terms of Cartesian unit vectors, and then expressing them again as spherical coordinates. By doing so, we see that
\begin{equation}
	\begin{split}
		\J &= \rv\times \h -\mu\rhat+\beta\mu(\nhat\cdot\rhat)^{2}(2\nhat-\rhat(\nhat\cdot\rhat))\\
		&+\beta\mu n_{r}^{2}\bigg(\rhat\int_{\theta_{0}}^{\theta}\d\eta\  n_{\theta}(\theta)\sin(\theta-\eta)+\thetahat\int_{\theta_{0}}^{\theta}\d\eta\  n_{\theta}\cos(\theta-\eta)\\
		&+\phihat\int_{\theta_{0}}^{\theta}\d\eta\  n_{\varphi}(\eta)G(\eta)\bigg).
	\end{split}
\end{equation}
Performing a dot product on $\J$ with the position vector $\r$, and then solving for $r$, we obtain the orbit equation
\begin{equation}
	\label{eq:this_eqn_here}
	r(\theta)=\frac{h^{2}}{\mu-\beta\mu n_{r}^{3}-\beta\mu F(\theta)+J\cos(\theta-\theta_{0})}
\end{equation}
where
\begin{equation*}
	F(\theta)=\int_{\theta_{0}}^{\theta}\d\eta\  n_{\theta}(\theta)\sin(\theta-\eta).
\end{equation*}
Now that we have two versions of the orbit equation, one from the constraint equation and another from \eqref{eq:orb_eqn_first_part}, we need to make sure that these equations are consistent with each other. This can be done by choosing the correct form of $\h$ so the two equations match. To obtain the correct $\h$, we differentiate the orbit equation with respect to the polar angle and set it to be proportional to $G(\theta)$, as the constraint equation \eqref{eq:constraint} suggests. It turns out that the correct squared magnitude of the specific angular momentum is given by
\begin{equation}
	h^{2}=h_{0}^{2}\exp\left(\int_{\theta_{0}}^{\theta}G(\nu) \d\nu\right)\left(\frac{\mu-\beta\mu n_{r}^{3}-\beta\mu F(\theta)+J\cos(\theta-\theta_{0})}{\mu-\beta\mu n_{r}^{3}+J}\right).
\end{equation}
Thus, substituting the correct form of specific angular momentum to equation \eqref{eq:this_eqn_here}, the orbit equation simplifies as
\begin{equation}
	\label{eq:radial_orbit_eqn}
	r(\theta)=\frac{h_{0}^{2}}{\mu-\beta\mu n_{r}^{3}+J}\exp\left(\int_{\theta_{0}}^{\theta}G(\nu)\d\nu\right).
\end{equation}
From the angular momentum function, we also determined a relationship among the initial conditions:
\begin{equation}
	\label{eq:initial_cond}
	r_{0}=\frac{h_{0}^{2}}{\mu-\beta\mu n_{r}^{2}+J}.
\end{equation}
The radial orbit equation we obtained depends on the lightness number, the initial angular momentum, and the nature of the constraint. By specifying the initial position of the sail and the sail's lightness number, we can determine the surface where the sail is expected to move.  However, since this orbit equation only involves the radial and polar coordinates, we need to determine another orbit equation that relates the azimuthal angle and the polar angle.

\subsection{Azimuthal Orbit Equation}
To obtain the azimuthal orbit equation we perform a dot product of the LRL vector, equation \eqref{eq:LRL_vector}, with $\thetahat/r^{2}$ and $\phihat/r^{2}$. This gives us
\begin{equation}
	\rdot\phidot\sin\theta=\frac{\beta\mu}{r^{2}}n_{r}\left[2n_{\varphi}(\theta)+\int_{\theta_{0}}^{\theta}\d\eta\ n_{\varphi}(\eta)G(\eta)\right]
\end{equation}
and
\begin{equation}
	\label{eq:rdot_thetadot}
	\rdot\thetadot=\frac{\beta\mu}{r^{2}}n_{r}^{2}\left(2n_{\theta}(\theta)+\int_{\theta_{0}}^{\theta}\d\eta\ n_{\theta}(\eta)G(\eta)\cos(\theta-\eta)\right).
\end{equation}
Dividing the two equations and integrating, we obtain the azimuthal orbit equation
\begin{equation}
	\label{eq:azimuthal_orbit_eqn}
	\varphi(\theta)=\varphi(\theta_{0})+\int_{\theta_{0}}^{\theta}\d\nu\frac{2n_{\varphi}(\nu)+A(\nu)}{\sin\nu [2n_{\theta}(\nu)+B(\nu)]}
\end{equation}
where $A$ and $B$ are given by
\begin{equation}
	\begin{split}
		A(\nu)&=\int_{\theta_{0}}^{\nu}\d\eta\ n_{\varphi}(\eta)\ G(\eta),\\
		B(\nu)&=\int_{\theta_{0}}^{\nu}\d\eta\ n_{\theta}(\eta)G(\eta)\cos(\nu-\eta).
	\end{split}
\end{equation}
Equations \eqref{eq:radial_orbit_eqn} and \eqref{eq:azimuthal_orbit_eqn} completely define an orbit in space. As said in the previous section, the radial orbit equation specifies the region or surface where the trajectory is expected to be located. On the other hand, the azimuthal orbit equation specifies the arc in space that the spacecraft will follow.  One can observe that the radial orbit equation is only dependent on the cone angle, whereas the azimuthal orbit equation, and hence the shape of the trajectory, depends only on the clock angle. Since the azimuthal orbit equation is independent of the lightness number, we can say that the shape of the orbit is scale-independent relative to $\beta$.

\subsection{The time derivatives $\rdot$, $\thetadot$, and $\phidot$}
From the LRL vector and the constraint equation, we can also obtain closed-form solutions of the radial, polar and azimuthal time derivatives. Substituting the constraint equation \eqref{eq:constraint} to equation \eqref{eq:rdot_thetadot} gives us 
\begin{equation}
	\Scale[0.95]{\displaystyle{r\thetadot^{2}G(\theta)=\frac{\beta\mu}{r^{2}}n_{r}^{2}\left[2n_{\theta}(\theta)+\int_{\theta_{0}}^{\theta}\d\eta\ n_{\theta}(\theta)G(\theta)\cos(\theta-\eta)\right]}}.
\end{equation}
Thus, the time rate of change of the polar angle is given by
\begin{equation}
	\label{eq:thetadot}
	\displaystyle{\thetadot=\left[\frac{\beta\mu n_{r}^{2}}{r^{3}G(\theta)}\left(2n_{\theta}(\theta)+\int_{\theta_{0}}^{\theta}\d\eta\ n_{\theta}(\eta)G(\eta)\cos(\theta-\eta)\right)\right]^{1/2}}.
\end{equation}
The radial component of the velocity $\rdot$ can be obtained from the constraint equation and our expression for $\thetadot$. Hence,
\begin{equation}
	\label{eq:rdot}
	\rdot=\left(\frac{\beta\mu}{r}n_{r}^{2}G(\theta)\left[2n_{\theta}(\theta)+\int_{\theta_{0}}^{\theta}\d\eta\ n_{\theta}(\eta)G(\eta)\cos(\theta-\eta)\right]\right)^{1/2}.
\end{equation}
Finally, the azimuthal time derivative $\phidot$ can be obtained from $\thetadot$ and $\d\varphi/\d\theta$. By chain rule,
\begin{equation}
	\label{eq:phidot}
	\phidot=\frac{\d\varphi}{\d\theta}\frac{\d\theta}{\d t}=\left[\frac{\beta\mu n_{r}^{2}}{r^{3}G(\theta)}\right]^{1/2}\frac{2n_{\varphi}(\theta)+\int_{\theta_{0}}^{\theta}\d\eta \ n_{\varphi}(\eta)G(\eta)}{\sin\theta\left[2n_{\theta}(\theta)+\int_{\theta_{0}}^{\theta}n_{\theta}(\eta)G(\eta)\cos(\theta-\eta)\right]^{1/2}}.
\end{equation}

From equations \eqref{eq:thetadot}, \eqref{eq:rdot}, and \eqref{eq:phidot}, the initial components of the velocity are given by
\begin{equation}
	\thetadot_{0}=\sqrt{\frac{2\beta\mu n_{r}^{2}n_{\theta}(\theta_{0})}{r_{0}^{3}G(\theta_{0})}}; \quad  \rdot_{0}=\sqrt{\frac{2\beta\mu n_{r}^{2}n_{\theta}(\theta_{0})G(\theta_{0})}{r_{0}}}; \quad  \phidot_{0}=\frac{1}{\sin\theta}\sqrt{\frac{2\beta\mu n_{r}^{2}(n_{\varphi}(\theta_{0}))^{2}}{r_{0}^{3}G(\theta_{0})n_{\theta}(\theta_{0})}}
\end{equation}

\subsection{Complete orbit characterization}
\label{sec:appx_complete_orbit}

An important question that arises in the radial orbit equation is the uniqueness of $J$ and $h_{0}$. While we can specify an initial position, lightness number, and cone angle, we still have two unknown quantities, which are $h_{0}$ and $J$. We can in fact relate $J$ with other quantities using the azimuthal orbit equation.  It is more desirable to express $\h_{0}$ and $J$ in terms of initial conditions since they can be directly measured and controlled. We express $h_{0}^{2}$ in terms of the initial position and velocity of the sail. Defining $\h_{0}$, we see that
\begin{equation}
	h_{0}^{2}=|\r_{0}\times\rv_{0}|^{2}=r_{0}^{4}\left(\thetadot_{0}^{2}+\sin^{2}\theta_{0}\phidot_{0}^{2}\right).
\end{equation}
Substituting expressions for the initial conditions $\thetadot_{0}$ and $\phidot_{0}$, and then simplifying, we see that
\begin{equation}
	h_{0}^{2}=\frac{2\beta\mu r_{0}n_{r}^{2}}{G(\theta_{0})}\left(n_{\theta}(\theta_{0})+\frac{n_{\varphi}^{2}(\theta_{0})}{n_{\theta}(\theta_{0})}\right)=\frac{2\beta\mu r_{0}n_{r}^{2}}{G(\theta_{0})}\left(\frac{1-n_{r}^{2}}{n_{\theta}(\theta_{0})}\right)
\end{equation}
But, from equation \eqref{eq:initial_cond}, we can express $r_{0}$ in terms of other $h_{0}$ and $J$. Hence, we can solve for $J$ in terms of the other quantities. It turns out that the condition for $J$ is that
\begin{equation}
	\label{eq:lrl_simp}
	J=-\mu+\beta\mu n_{r}^{3}+\frac{2\beta\mu}{G(\theta_{0})}\left(\frac{n_{r}^{2}(1-n_{r}^{2})}{n_{\theta}(\theta_{0})}\right)\ge 0
\end{equation}
Hence, as long as the magnitude of the generalized LRL vector follows equation \eqref{eq:lrl_simp}, we are sure that the trajectory satisfies the equation of motion.  Given some initial conditions $\r_{0}$ and $\rv_{0}$, we can choose a set $\{J,h_{0},\nhat_{0}\}$ that will make the orbit equation consistent with the equation of motion. If one of these three dynamical quantities is set, then we can determine the other two uniquely. In practice, the dependence of $\nhat$ to the position of the sail can be set based on the mission objectives. Hence, the LRL vector and the initial angular momentum can then be obtained from $\nhat$. 

If a trajectory does not satisfy this condition, the resulting orbit is extraneous and hence the obtained solution becomes inconsistent with the equation of motion.

\subsection{Summary}

In summary, our orbit determination scheme consists of the following parts:
\begin{itemize}
	\item Consider in the mission planning the initial conditions of the system.
	\item Choose a constraint equation \eqref{eq:constraint} based on the expected geometry of the surface. 
	\item From equation \eqref{eq:radial_orbit_eqn}, the constraint equation immediately gives the radial orbit equation by specifying an initial position $(r_{0},\theta_{0},\varphi_{0})$ and velocity $(\rdot_{0},\thetadot_{0},\phidot_{0})$.
	\item Determine the azimuthal orbit equation using equation \eqref{eq:azimuthal_orbit_eqn}. In very special cases, $\varphi(\theta)$ can be obtained analytically. 
	\item Other physical quantities such as velocities and angular momenta can be obtained. 
\end{itemize}

\section{Application to Orbits Constrained on Cylinders}
\label{sec:applns_cyl}
\begin{figure}[H]
	\centering
	\includegraphics[scale=0.35]{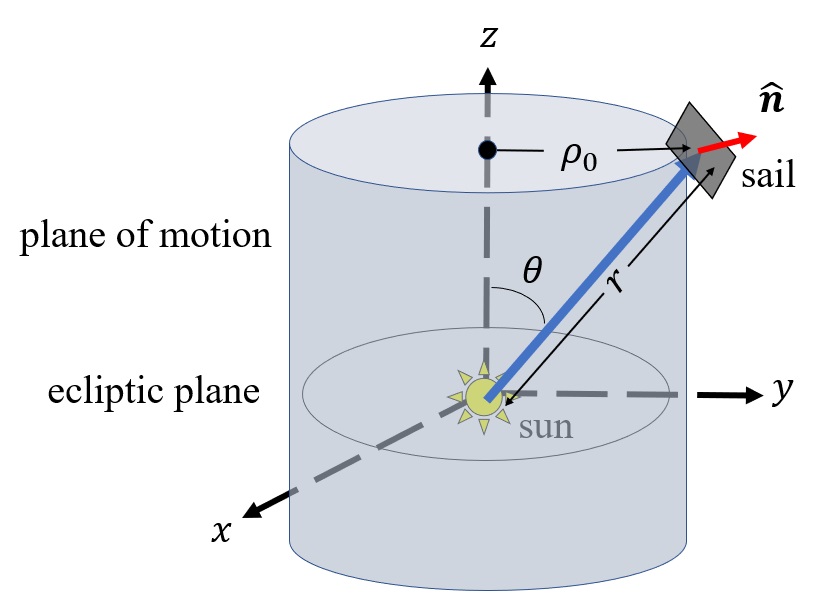}
	\caption{A sail's orbit constrained on a cylinder. The shaded region is the surface where the sail is constrained to move.}
	\label{fig:cyl_diagram}
\end{figure}

As an example, consider an orbit constrained on a cylinder (\textit{See Figure} \ref{fig:cyl_diagram}). For a cylinder, $\rho_{0}=r\sin\theta$, where $\rho_{0}$ is a constant. Differentiating $\rho_{0}$ with respect to time gives us a relationship between the radial and polar components of the velocities:
\begin{equation}
	\rdot=rG(\theta)\thetadot=-r\thetadot\cot\theta.
\end{equation}
Comparing with equation \eqref{eq:constraint}, we choose $G(\theta)=-\cot\theta$. The radial orbit equation is given by
\begin{equation}
	r(\theta)=r_{0}\frac{\csc\theta}{\csc\theta_{0}}
\end{equation}
where the initial position $r_{0}$ is specified by equation \eqref{eq:initial_cond}. The azimuthal orbit equation would however depend on the functional dependence of the polar and azimuthal components of the unit normal vector or equivalently on how the clock angle changes through $\theta$.

\subsection{Constant Cone and Clock Angles}
\begin{figure*}[h!]
	\centering
	\begin{subfigure}{0.5\textwidth}
		\centering
		\includegraphics[scale=0.25]{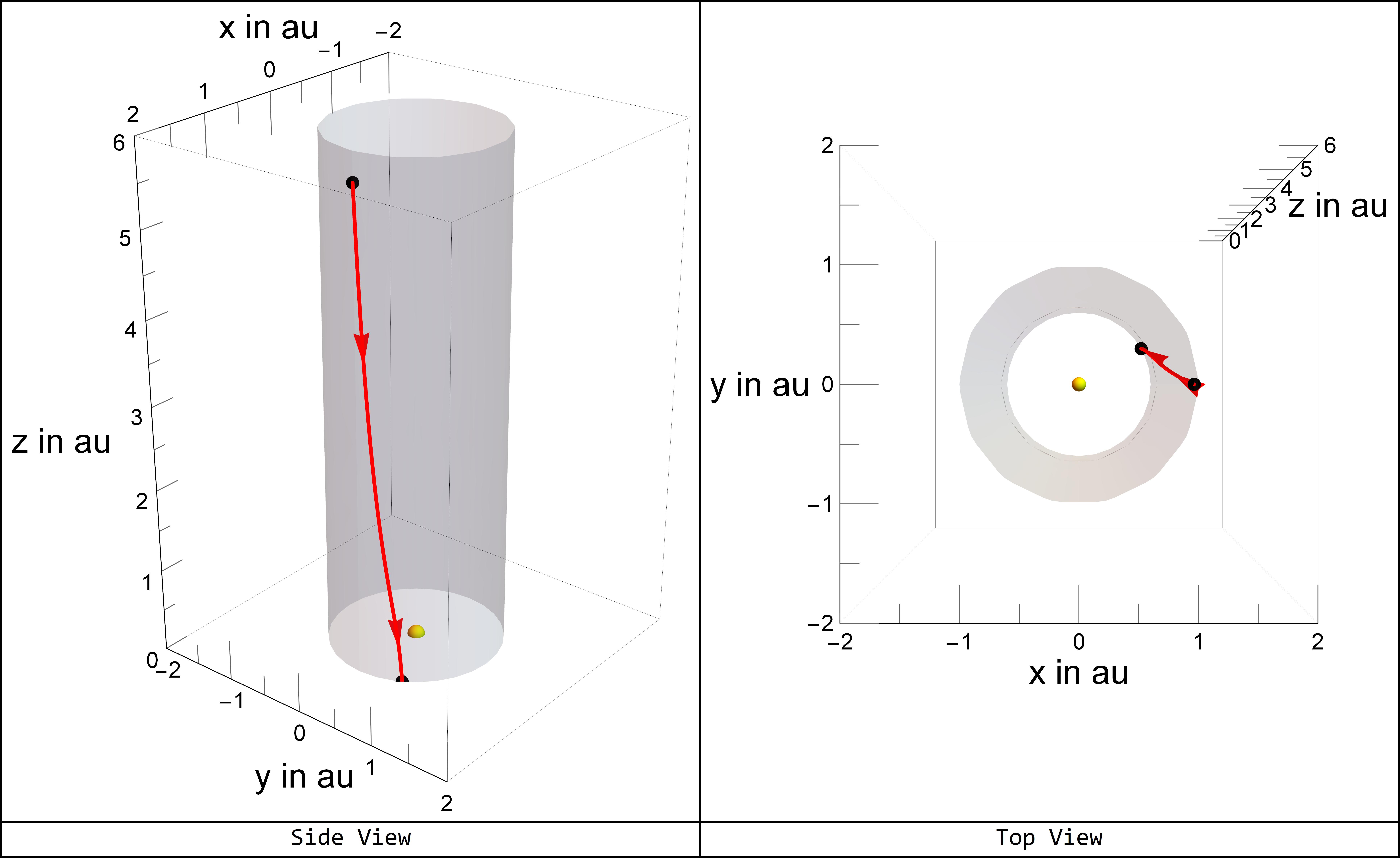}
		\caption{$\delta=15^{\circ}$}
		\label{fig:delta_15}
	\end{subfigure}%
	\begin{subfigure}{0.5\textwidth}
		\centering
		\includegraphics[scale=0.25]{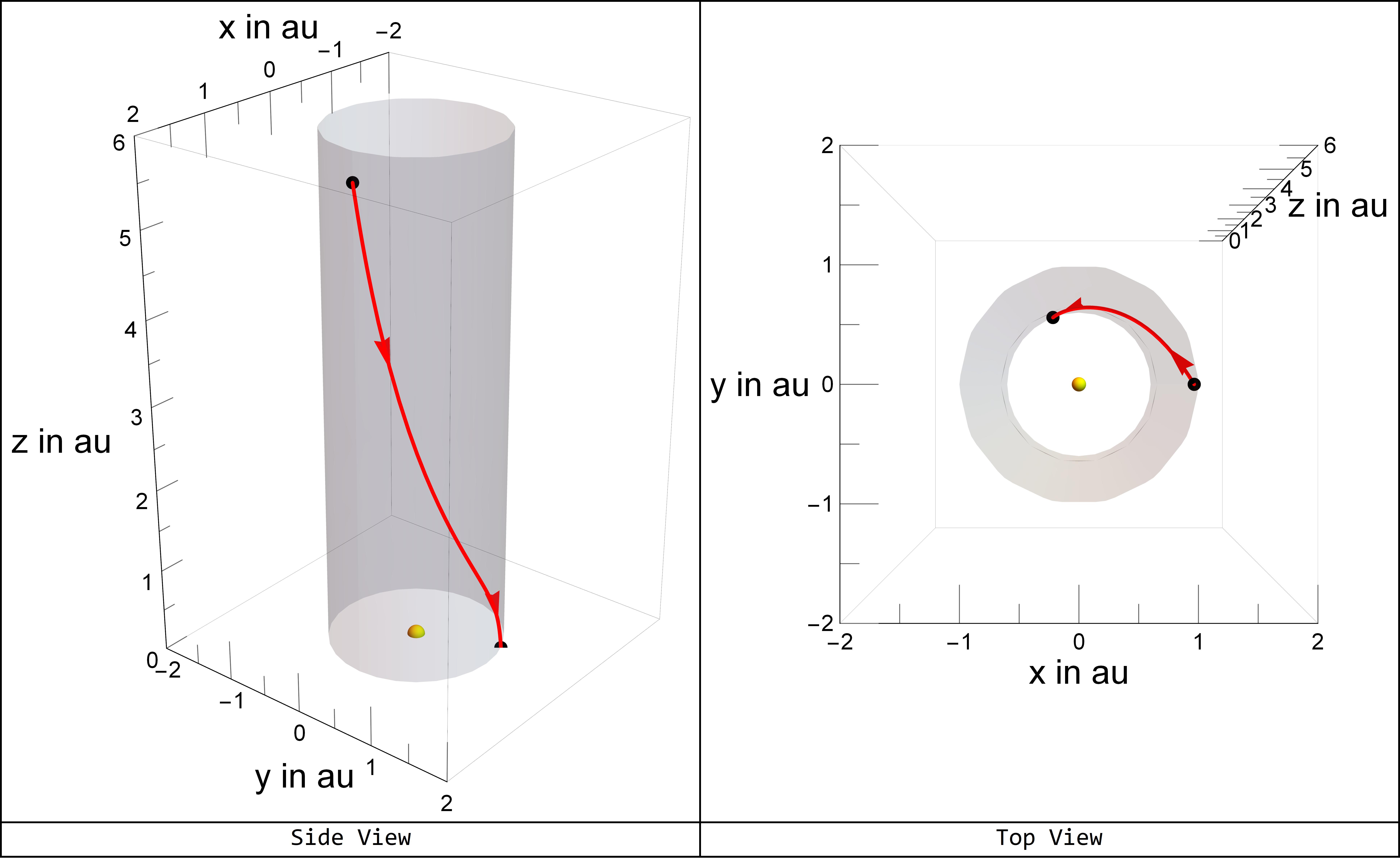}
		\caption{$\delta=45^{\circ}$}
		\label{fig:delta_30}
	\end{subfigure}
	\begin{subfigure}{0.5\textwidth}
		\centering
		\includegraphics[scale=0.25]{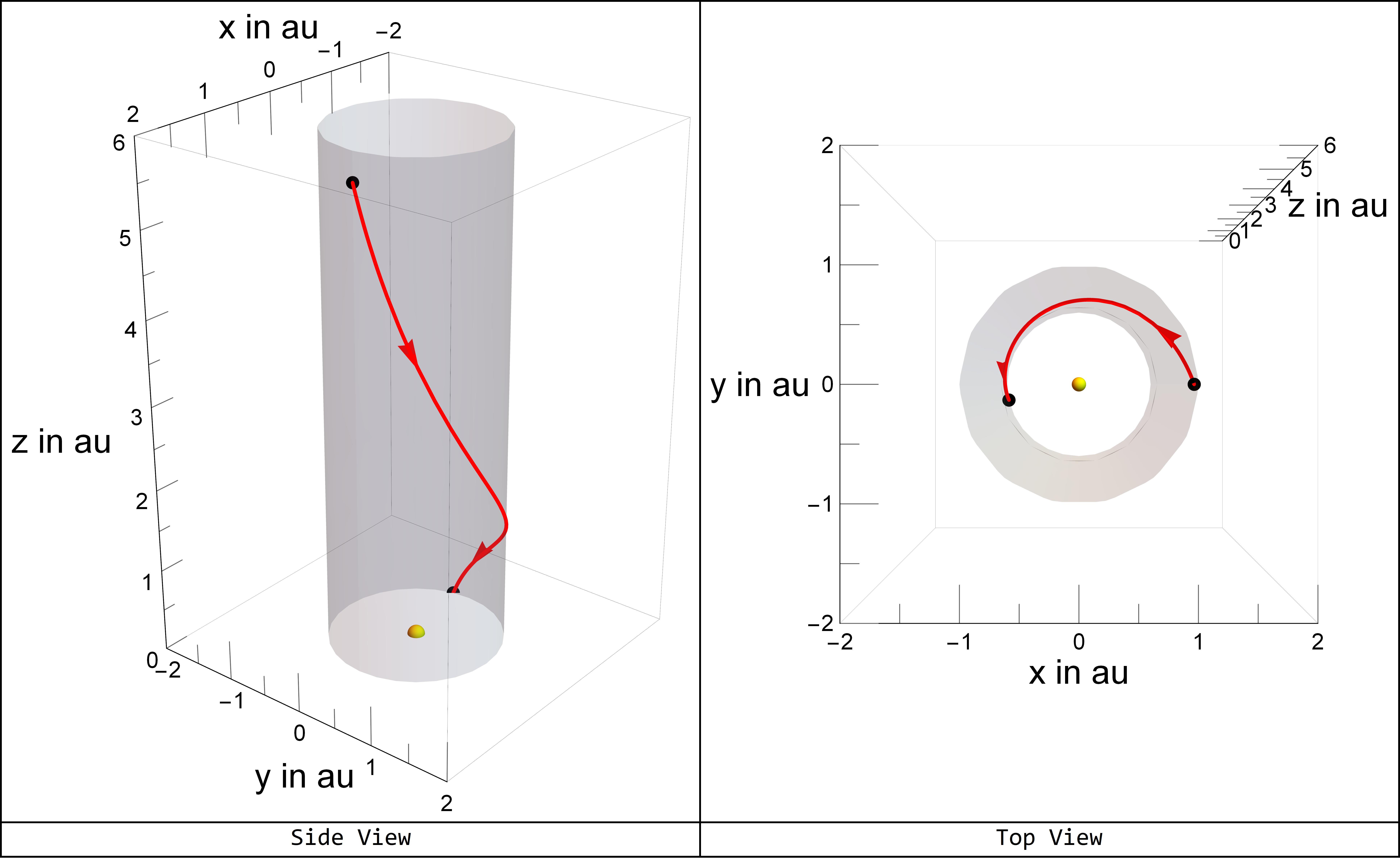}
		\caption{$\delta=60^{\circ}$}
		\label{fig:delta_60}
	\end{subfigure}%
	\begin{subfigure}{0.5\textwidth}
		\centering
		\includegraphics[scale=0.25]{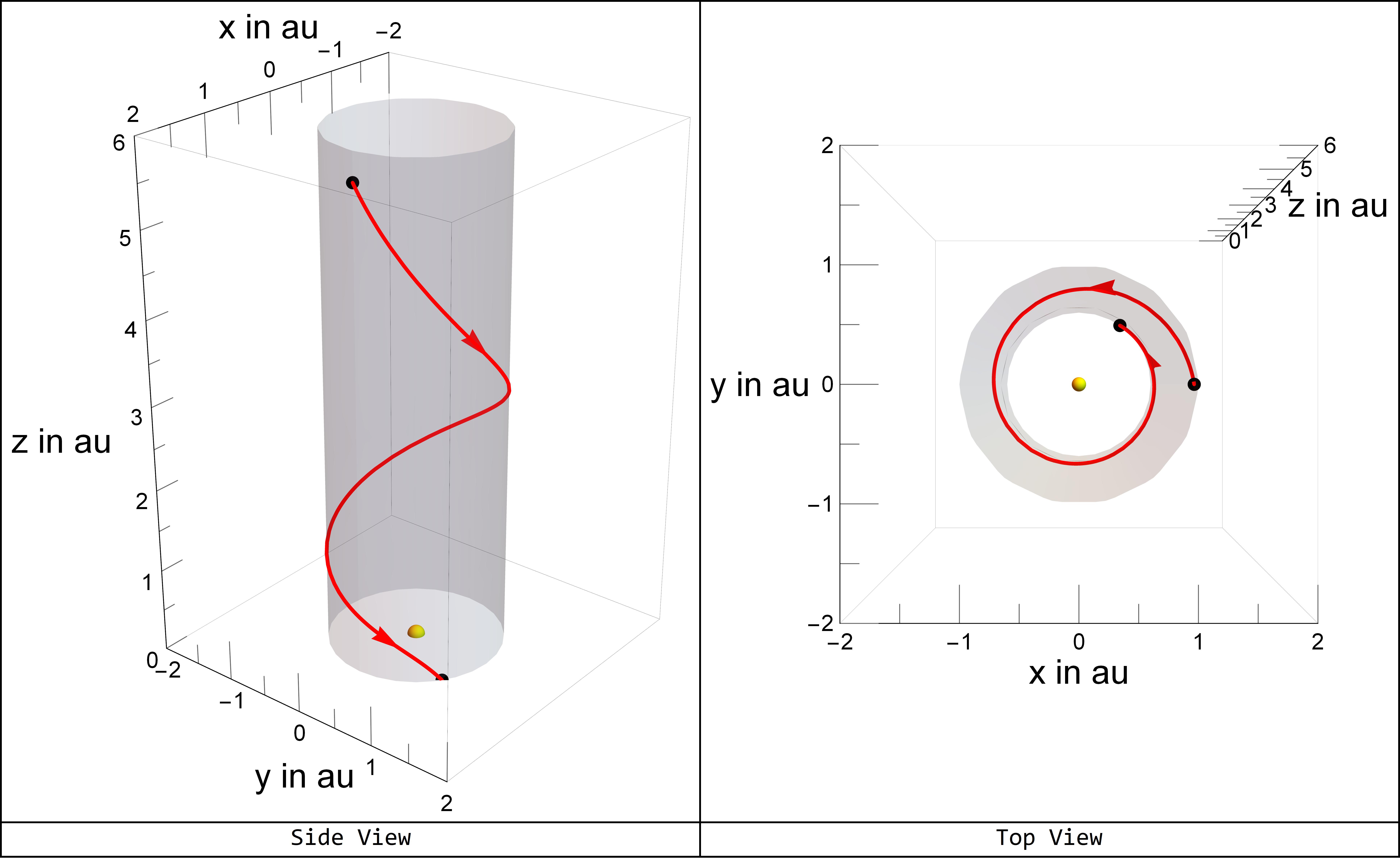}
		\caption{$\delta=75^{\circ}$}
		\label{fig:delta_75}
	\end{subfigure}
	\caption{Orbits constrained on a cylinder. We set the initial radial position to be at $r_{0}=1\ 
		\text{au}$. The initial cone angle is set at $\alpha=-45^{\circ}$. The sun is specified to be at the origin of the coordinate system.}
	\label{fig:cyl_orbits}
\end{figure*}

We first consider the case when both the cone and clock angles are constant throughout the mission such that if $\delta$ is the clock angle, the azimuthal orbit equation becomes
\begin{equation}
	\label{eq:cyl_azm_orb_eqn_const_angles}
	\begin{split}
		\varphi(\theta)&=\varphi(\theta_{0})+\displaystyle{\tan\delta\int_{\theta_{0}}^{\theta}\d\nu \frac{2-\ln\left(\frac{\sin\nu}{\sin\theta_{0}}\right)}{\left[1+\cos(\nu-\theta_{0})-\ln\left(\frac{\tan(\nu/2)}{\tan(\theta_{0}/2)}\right)\cos\nu\right]\sin\nu}}.
	\end{split}
\end{equation}
The integral can be evaluated numerically, giving us the the trajectory of the sail at different values of the clock angle. Shown in Figure \ref{fig:cyl_orbits} are the orbits constrained on a cylindrical surface at different values of $\delta$. For small $\delta$, the azimuthal position hardly changes. The component of the torque in the azimuthal direction increases as the clock angle increases. This implies an increase in the azimuthal acceleration, speeding up the spacecraft in the $\phihat$ direction. Consequently, more turns are observed as $\delta$ increases.

We can gain more understanding of what is happening at the sail's trajectory by looking at the radial, azimuthal, and polar components of the velocity as functions of the polar angle. One can show from equations \eqref{eq:thetadot}, \eqref{eq:rdot} and \eqref{eq:phidot} that if
\begin{equation}
	b=1+\cos(\theta-\theta_{0})-\cos\theta\ln\left(\frac{\tan(\theta/2)}{\tan(\theta_{0}/2)}\right),
\end{equation}
the time rate of change of radial, polar, and azimuthal coordinates of the spacecraft are given by equations \eqref{eq:cyl_polar_velocity}, \eqref{eq:cyl_azimuthal_velocity} and \eqref{eq:cyl_radial_velocity}:
\begin{equation}
	\label{eq:cyl_polar_velocity}
	\displaystyle{\frac{\d\theta}{\d t}= \left[\frac{-\beta\mu  n_{r}^{2}n_{\theta}}{r^{3}\cot\theta}b\right]^{1/2}}
\end{equation}
\begin{equation}
	\label{eq:cyl_azimuthal_velocity}
	\frac{\d\varphi}{\d t}= \left[\frac{-\beta\mu}{r^{3}\cot\theta}\frac{n_{r}^{2}n_{\varphi}^{2}}{n_{\theta}}\frac{1}{b}\right]^{1/2}\frac{2-\ln\left[\frac{\sin\theta}{\sin\theta_{0}}\right]}{\sin\theta}
\end{equation}
\begin{equation}
	\label{eq:cyl_radial_velocity}
	\frac{\d r}{\d t}=\left(\frac{-\beta\mu}{r} n_{r}^{2}n_{\theta}b\cot\theta \right)^{1/2}.
\end{equation}
The range of allowable values for both $\alpha$ and $\delta$ is indicated by the velocity components. While the azimuthal orbit equation does not indicate any limitation on the value of $\alpha$, the velocities must be physical, hence, the cone angle and the clock angle have opposite signs.  A negative cone angle indicates that the spacecraft is approaching the sun, which is consistent from the observation that the radial distance $r$ becoming smaller in time. Thus, for the velocities to be defined, the clock angle should be positive. 

\begin{figure*}[h!]
	\centering
	\begin{subfigure}{1.0\textwidth}
		\centering
		\includegraphics[scale=0.5]{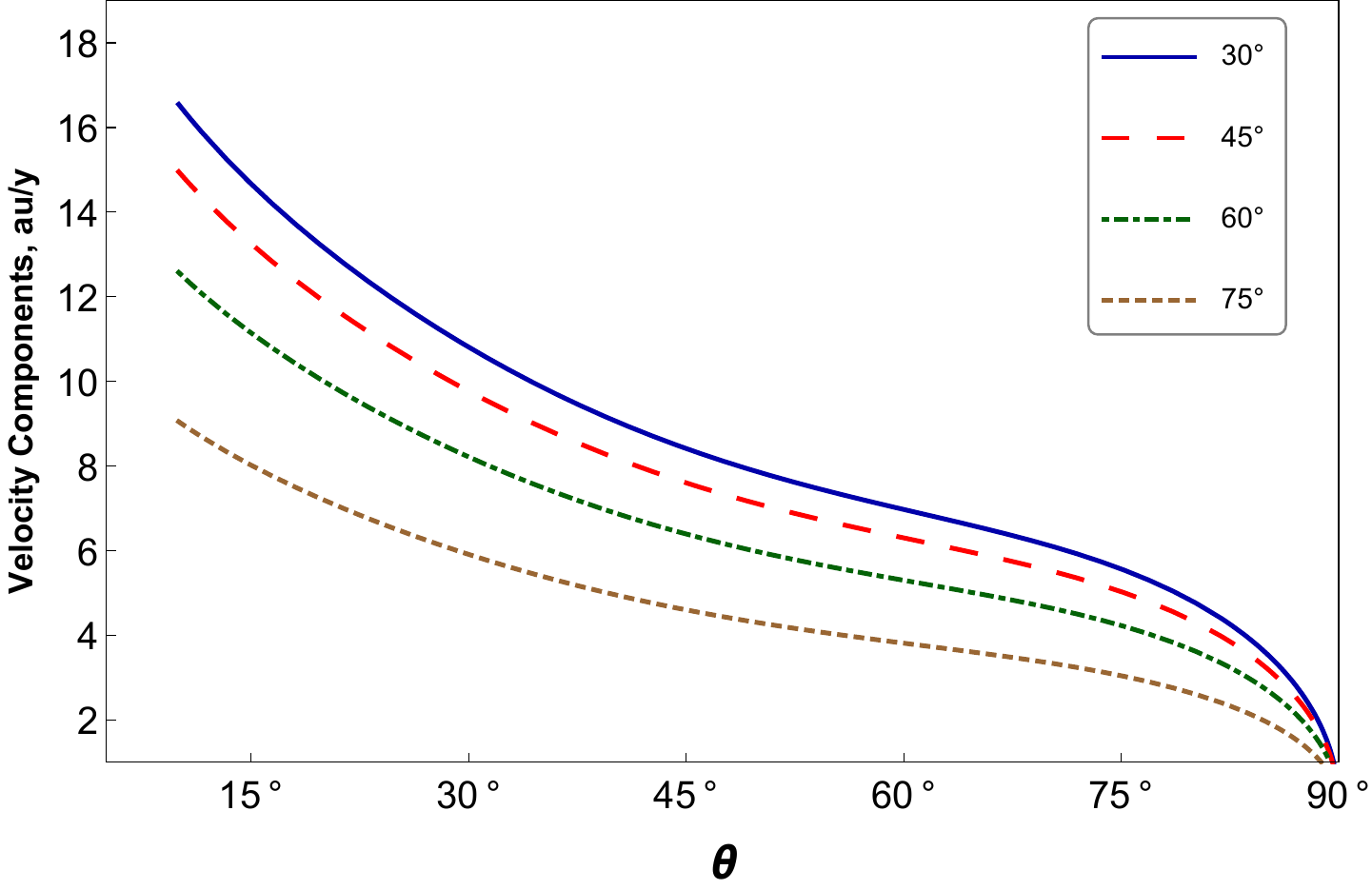}
		\caption{Radial component}
		\label{fig:vr_alpha=-45}
	\end{subfigure}
	\begin{subfigure}{1.0\textwidth}
		\centering
		\includegraphics[scale=0.5]{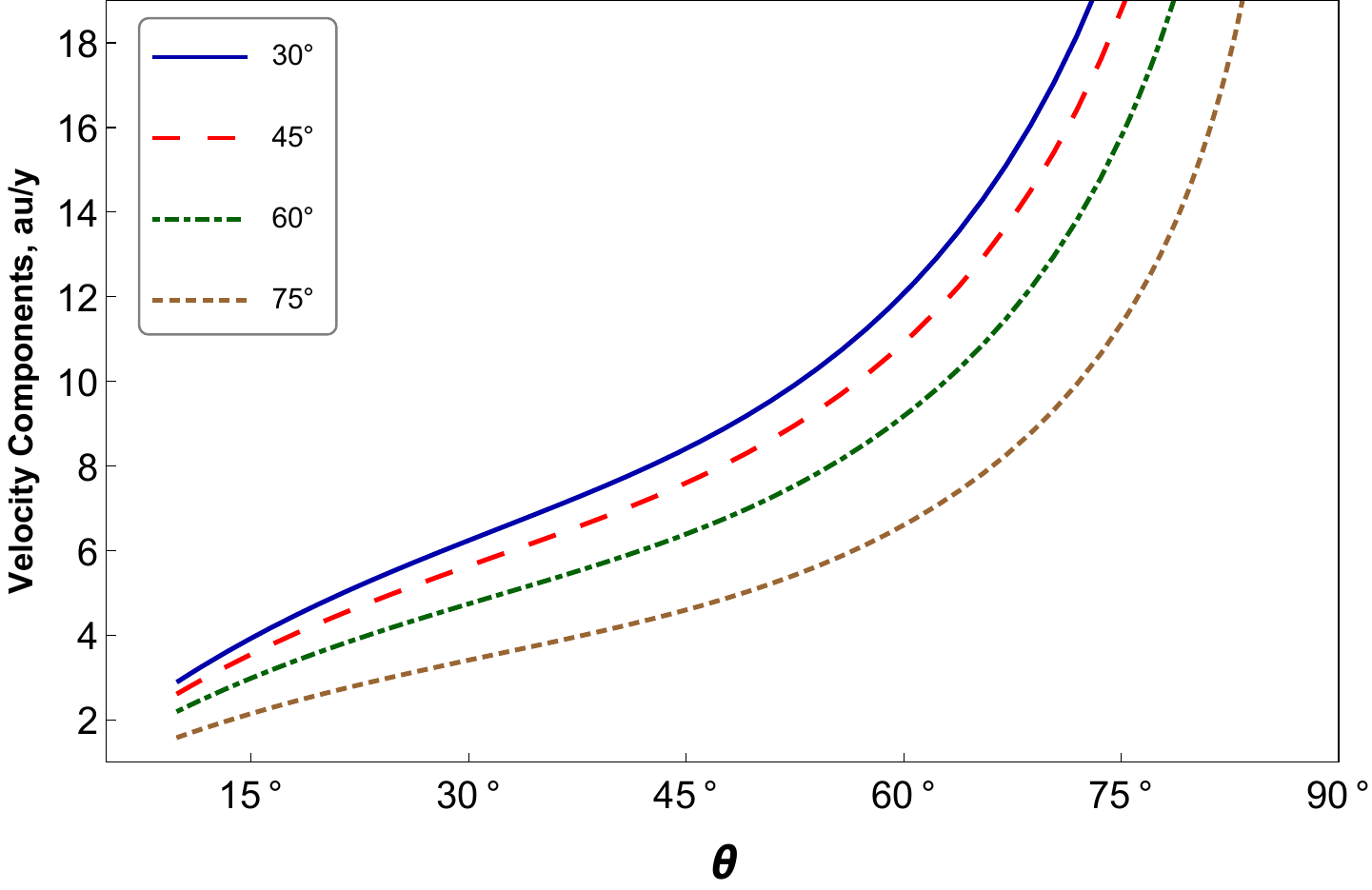}
		\caption{Polar component}
		\label{fig:vtheta_alpha=-45}
	\end{subfigure}
	\begin{subfigure}{1.0\textwidth}
		\centering
		\includegraphics[scale=0.5]{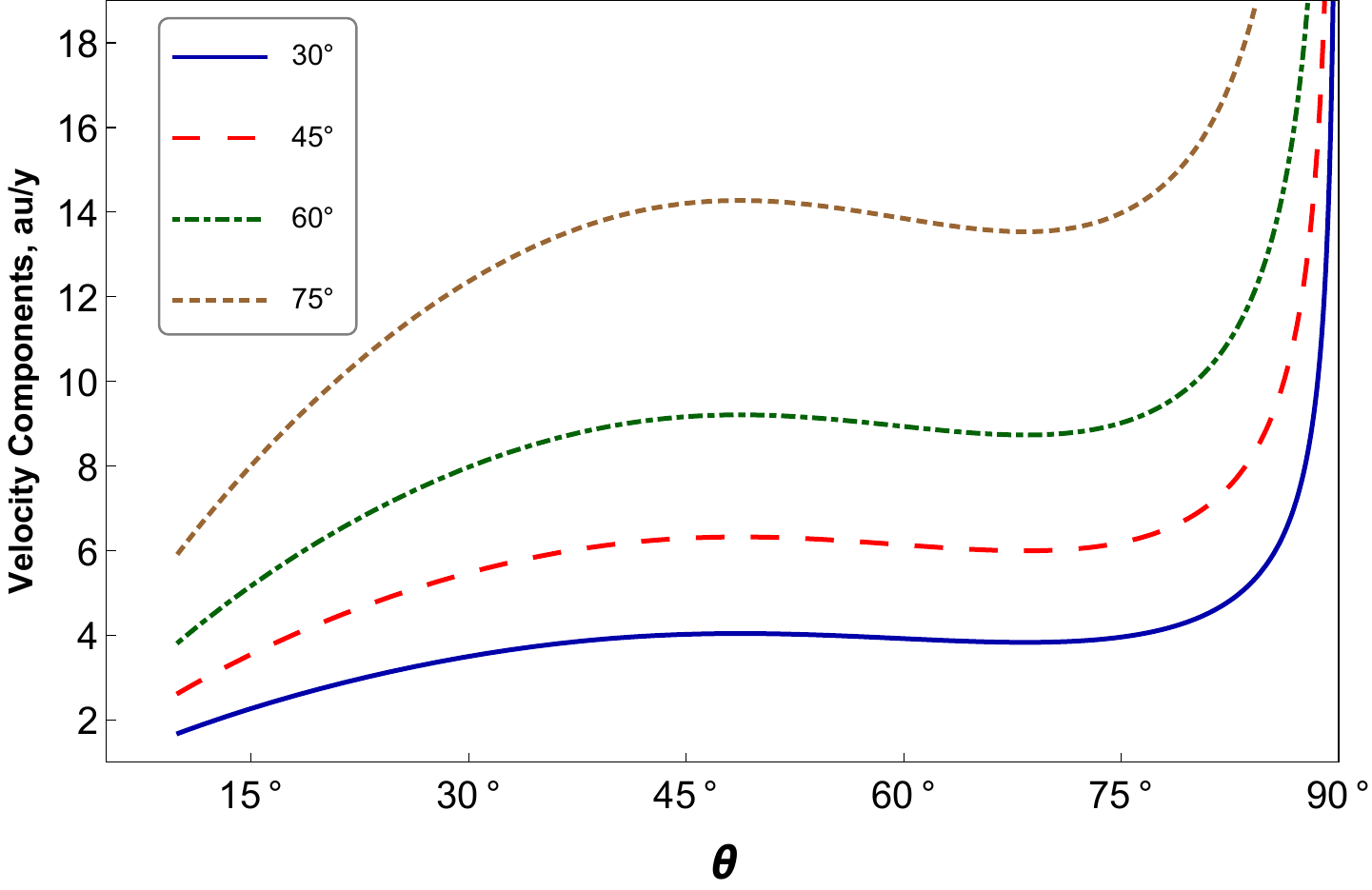}
		\caption{Azimuthal component}
		\label{fig:vphi_alpha=-45}
	\end{subfigure}
	\caption{The radial, polar, and azimuthal components of the velocity for different values of the clock angle $\delta$. We set the initial polar angle at $\theta_{0}=10^{\circ}$.}
	\label{fig:cyl_orbits_velocities}
\end{figure*}

From equations \eqref{eq:cyl_polar_velocity}, \eqref{eq:cyl_azimuthal_velocity}, and \eqref{eq:cyl_radial_velocity}, we can make a phase space plot of the velocity components as a function of $\theta$. (\textit{See Figure \ref{fig:cyl_orbits_velocities}}).  At small values of $\theta$, the radial component has a greater effect in the total velocity of the spacecraft than the polar and the azimuthal components. As $\theta$ increases, the polar component of the velocity increases until the spacecraft reaches $\theta=90^{\circ}$ for which the sail's radial velocity vanishes but its polar velocity is at the maximum. This can be explained by considering the forces acting on the sail. For the sail to maintain its orbit, the gravitational force must counteract the radial component of the solar radiation pressure. However, the trajectory of the sail will be constrained on a sphere if the strength of these two radial forces are equal. Since we want to make the spacecraft to be constrained on the cylinder, the gravitational force must be greater than the radial component of the solar radiation pressure so that the radial distance $r$ decreases through time. The inward net force in the radial direction is compensated by the polar component of the solar radiation pressure force, resulting in a net force that is downward for smaller values of $\delta$. Since the velocities involved are positive and non-zero, for smaller values of the clock angle, the direction of the trajectory is downward, consistent with what was shown in Figure \ref{fig:cyl_orbits}. 

The deflection in the over-all downward trajectory of the sail is due to the azimuthal component of the solar radiation pressure. For smaller values of $\delta$, this deflection is hardly noticeable. However, as $\delta$ increases, the component of the torque in the direction of $\phihat$ increases, causing an increase in this velocity component. This increase in the azimuthal component of the torque is also reflected in the $\tan\delta$ factor in the azimuthal orbit equation \eqref{eq:cyl_azm_orb_eqn_const_angles}. Consequently, the trajectory of the sail becomes more helical as the clock angle approaches $90^{\circ}$.

\newpage

\subsection{Periodic Components of the Unit Normal Vector}
\begin{figure*}[h!]
	\centering
	\begin{subfigure}{0.5\textwidth}
		\centering
		\includegraphics[scale=0.25]{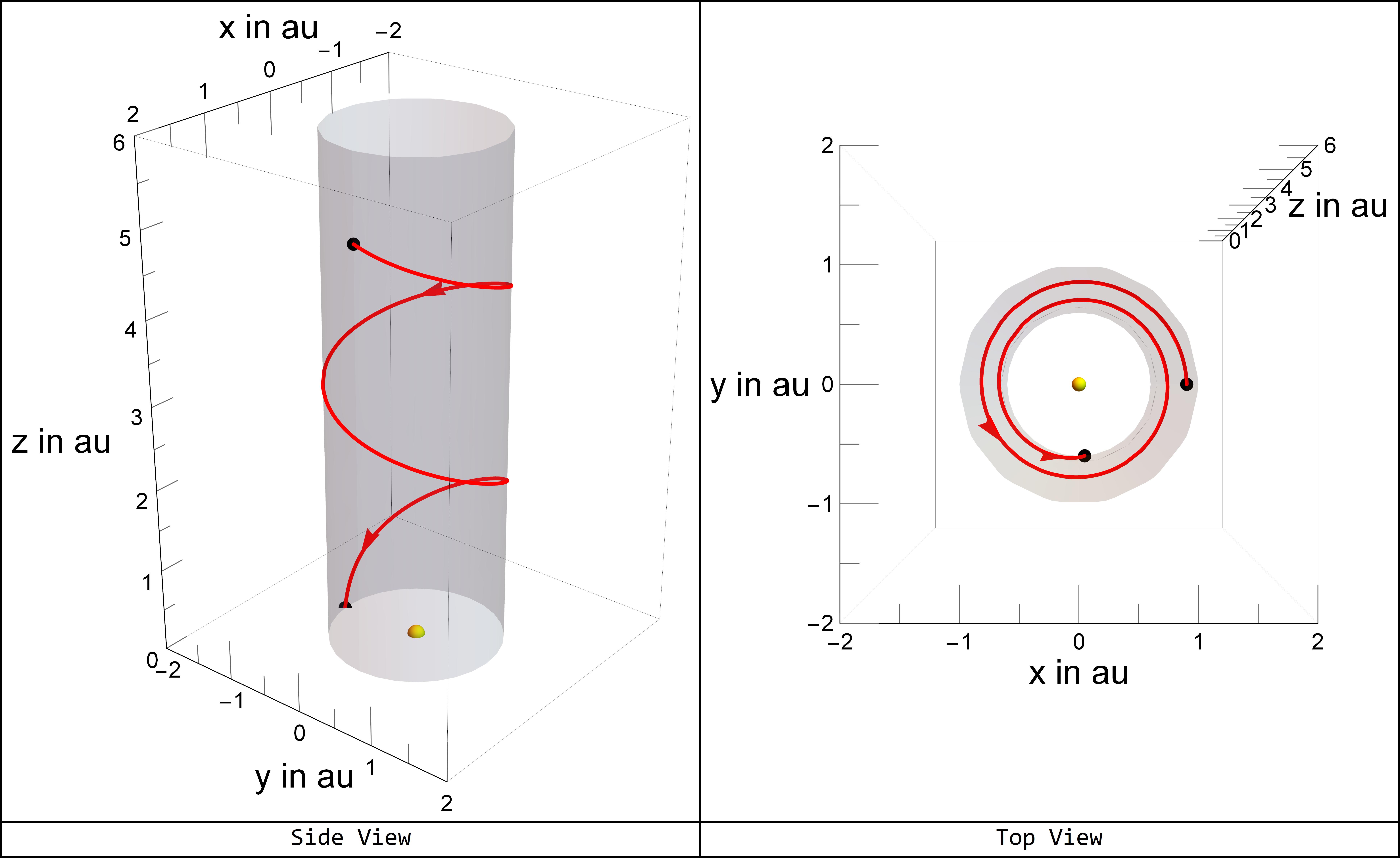}
		\caption{$k=1/3$}
		\label{fig:cyl_k=1-3}
	\end{subfigure}%
	\begin{subfigure}{0.5\textwidth}
		\centering
		\includegraphics[scale=0.25]{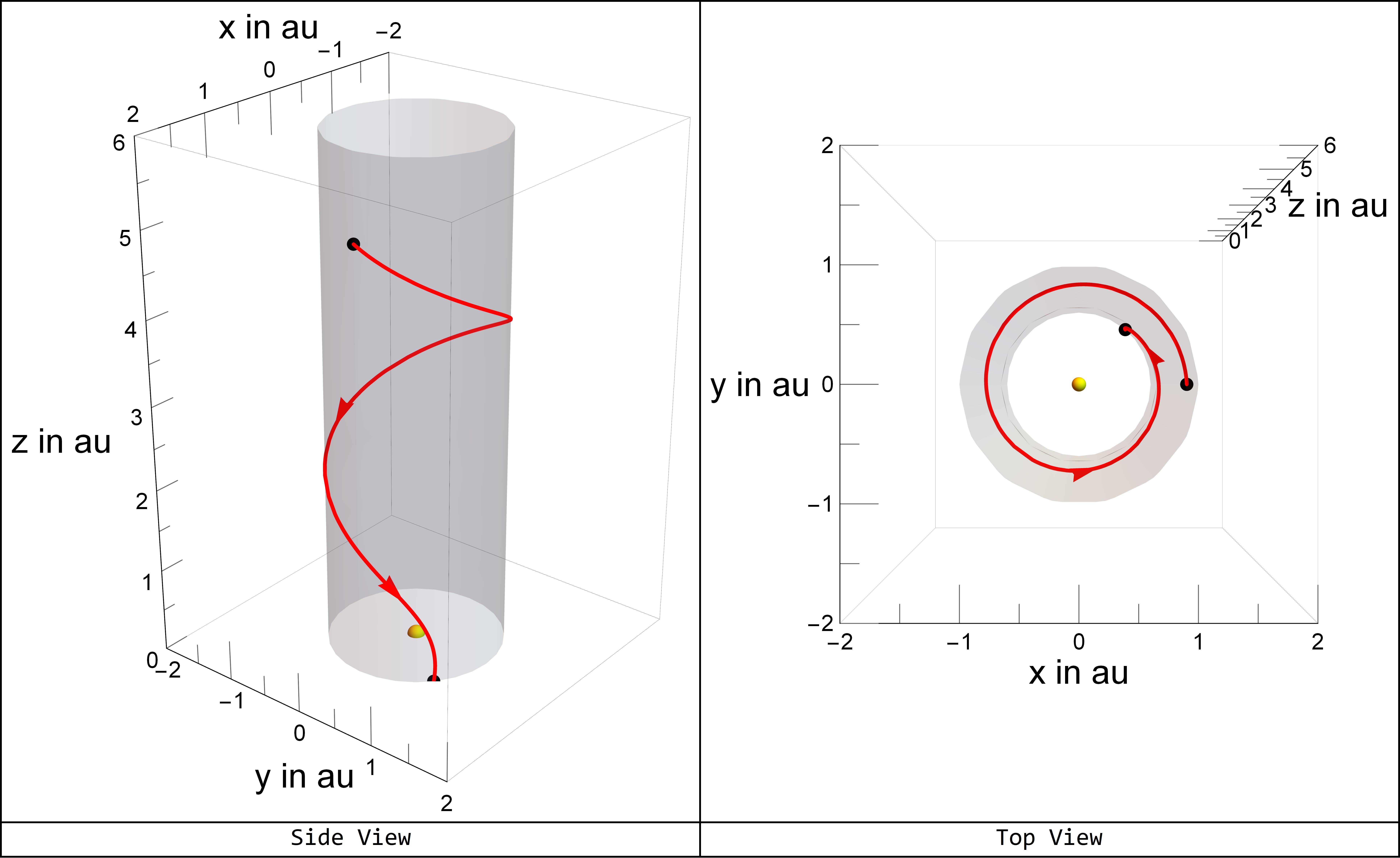}
		\caption{$k=1/2$}
		\label{fig:cyl_k=1-2}
	\end{subfigure}
	\begin{subfigure}{0.5\textwidth}
		\centering
		\includegraphics[scale=0.25]{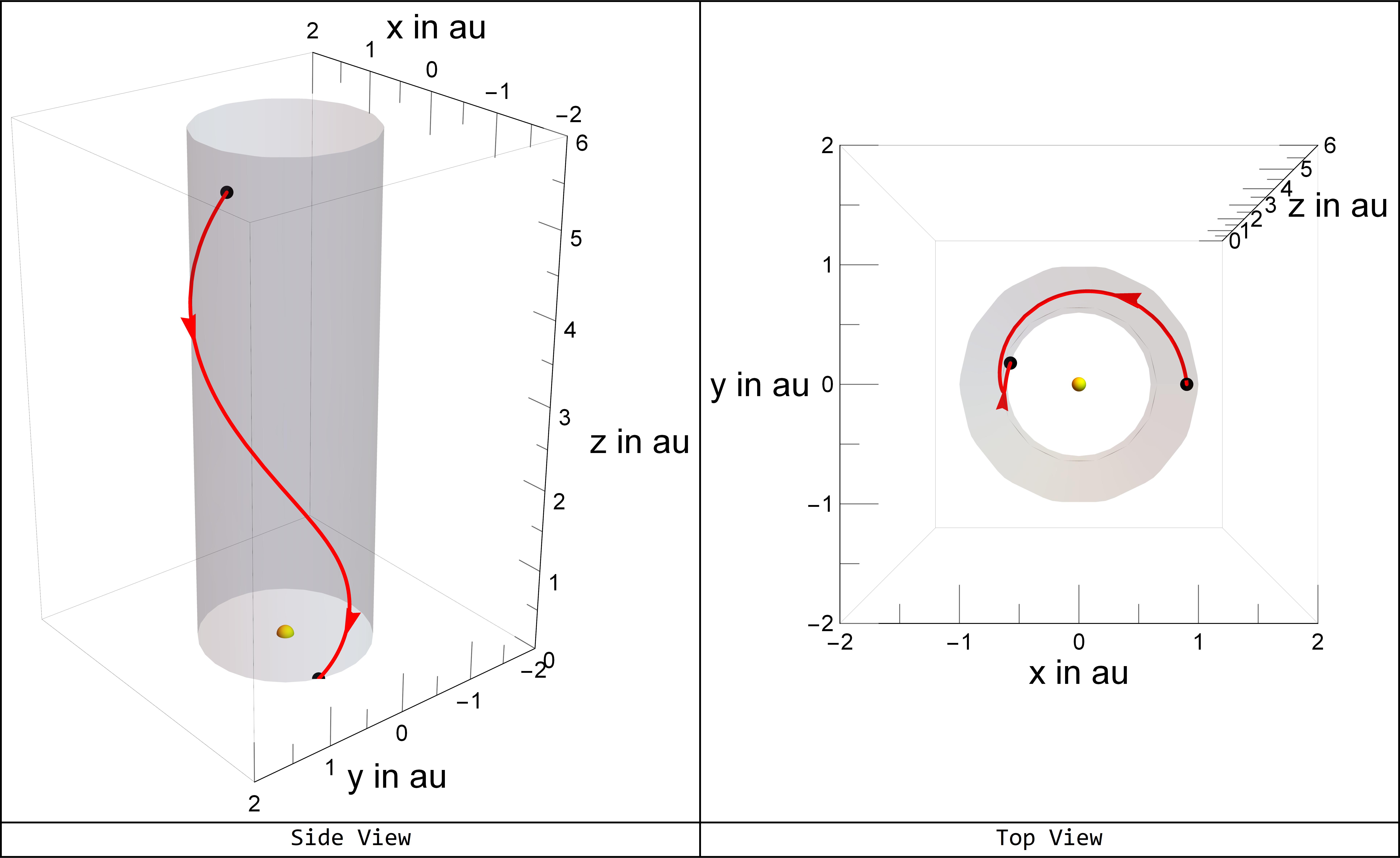}
		\caption{$k=1$}
		\label{fig:cyl_k=1-0}
	\end{subfigure}%
	\begin{subfigure}{0.5\textwidth}
		\centering
		\includegraphics[scale=0.25]{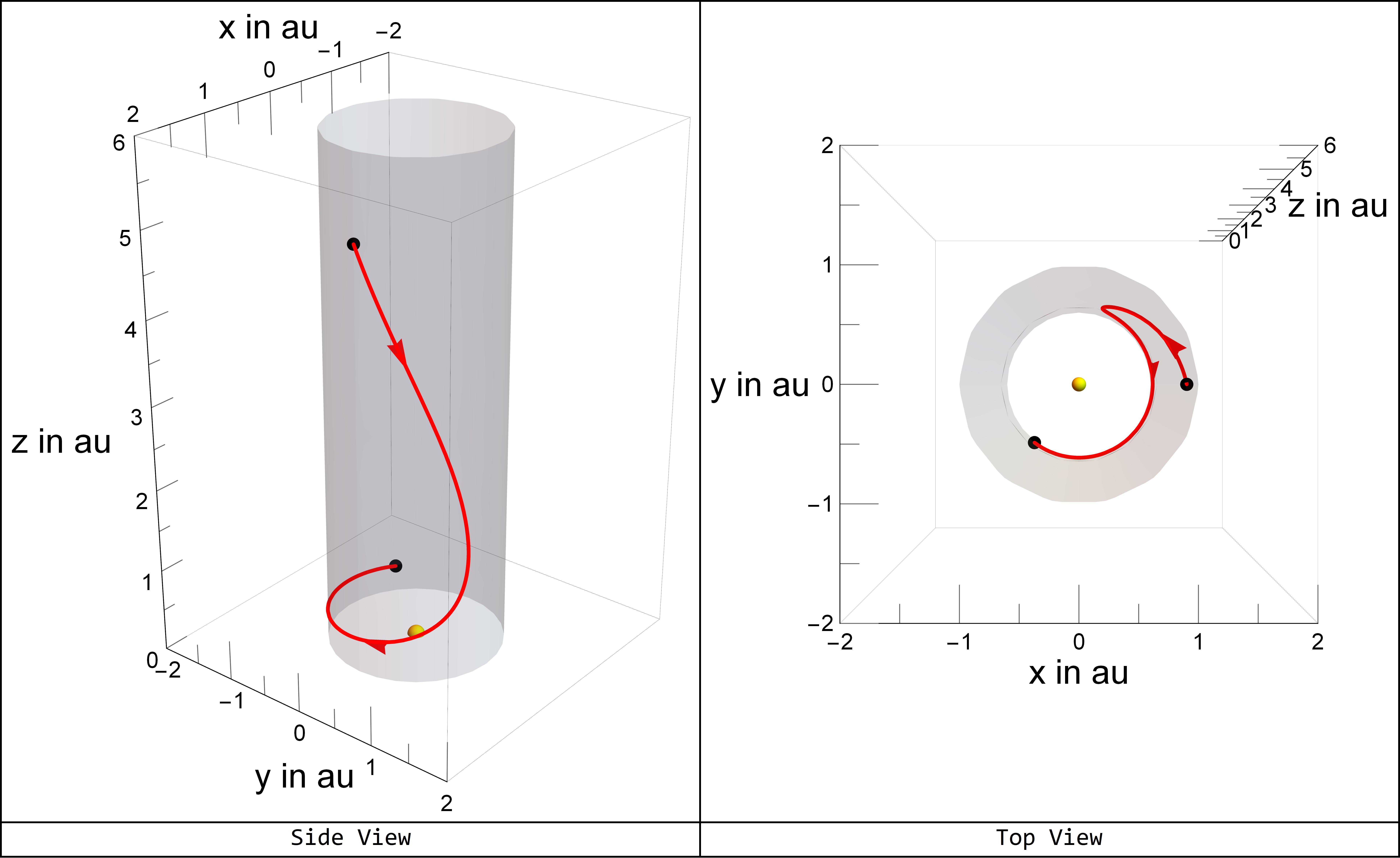}
		\caption{$k=2$}
		\label{fig:cyl_k=2-0}
	\end{subfigure}
	\caption{Orbits constrained on a cylinder for periodic $n_{\theta}$ and $n_{\varphi}$. We assume that the sail approaches the ecliptic plane at the earth-sun distance. The initial cone angle is set at $\alpha=-45^{\circ}$. The sun is specified to be at the origin of the coordinate system.}
	\label{fig:cyl_orbits_periodic_clock}
\end{figure*}

\begin{figure*}[h!]
	\centering
	\begin{subfigure}{1\textwidth}
		\centering
		\includegraphics[scale=0.45]{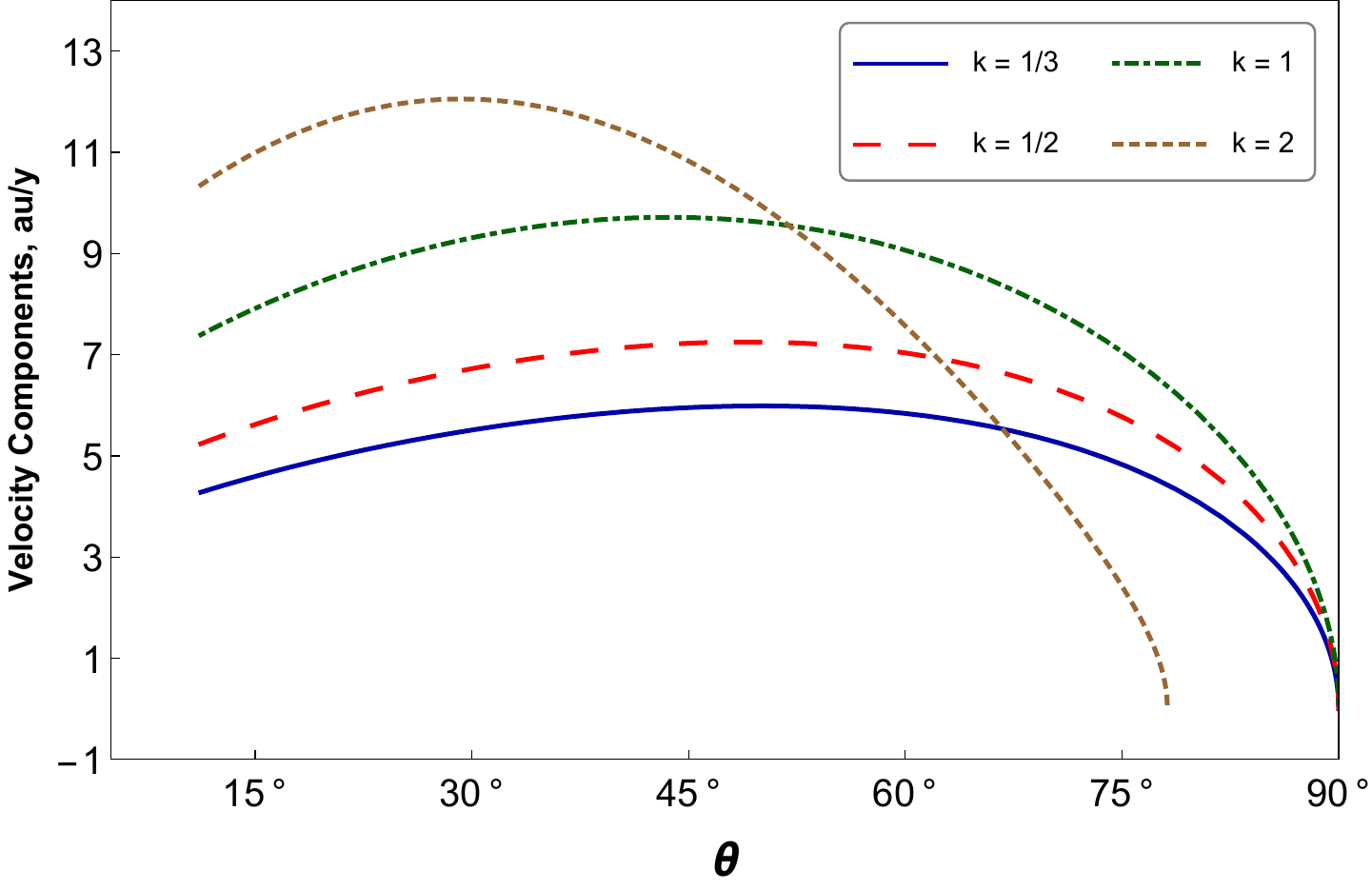}
		\caption{Radial Component}
		\label{fig:vr_k}
	\end{subfigure}
	\begin{subfigure}{1\textwidth}
		\centering
		\includegraphics[scale=0.45]{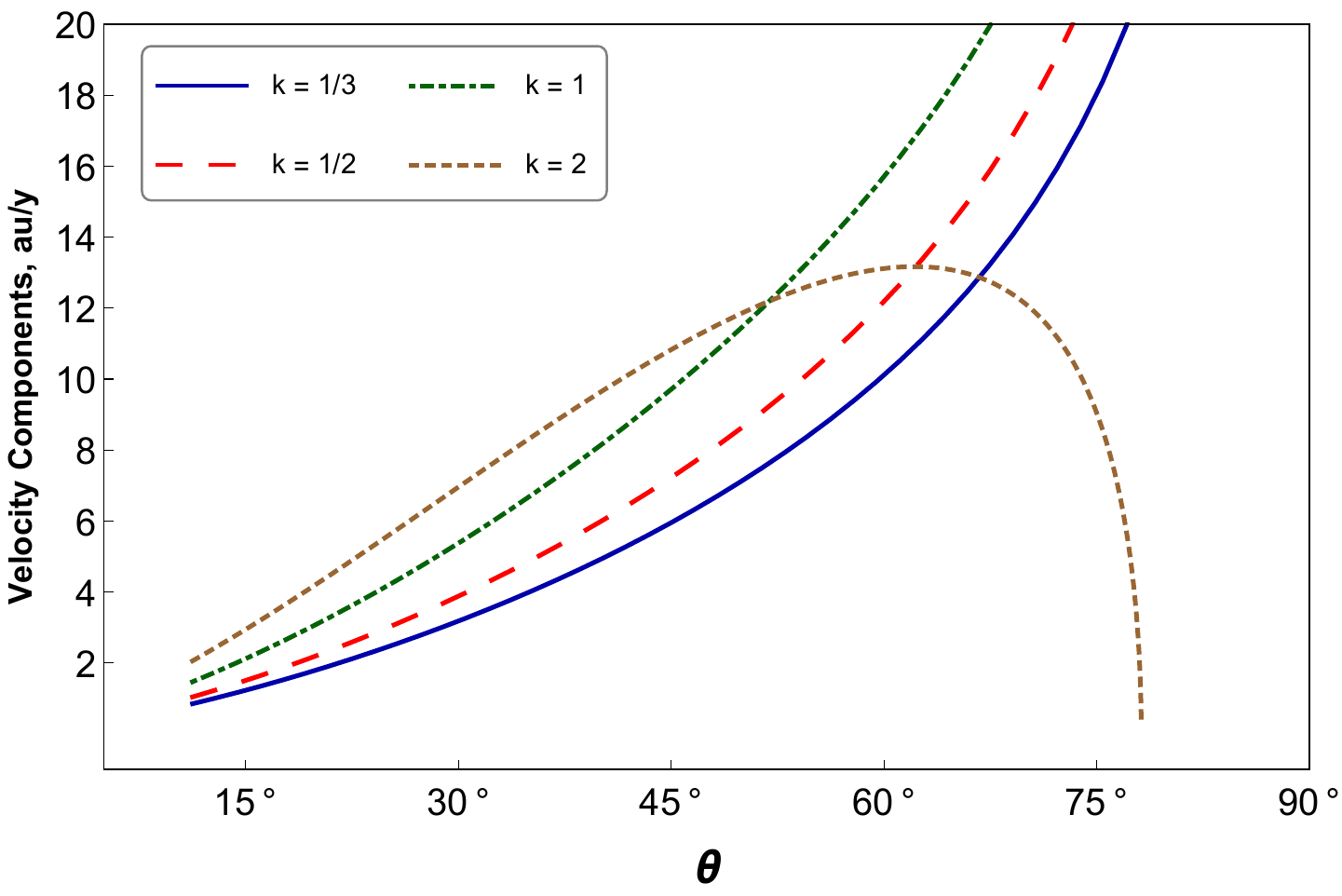}
		\caption{Polar Component}
		\label{fig:vtheta_k}
	\end{subfigure}
	\begin{subfigure}{1\textwidth}
		\centering
		\includegraphics[scale=0.45]{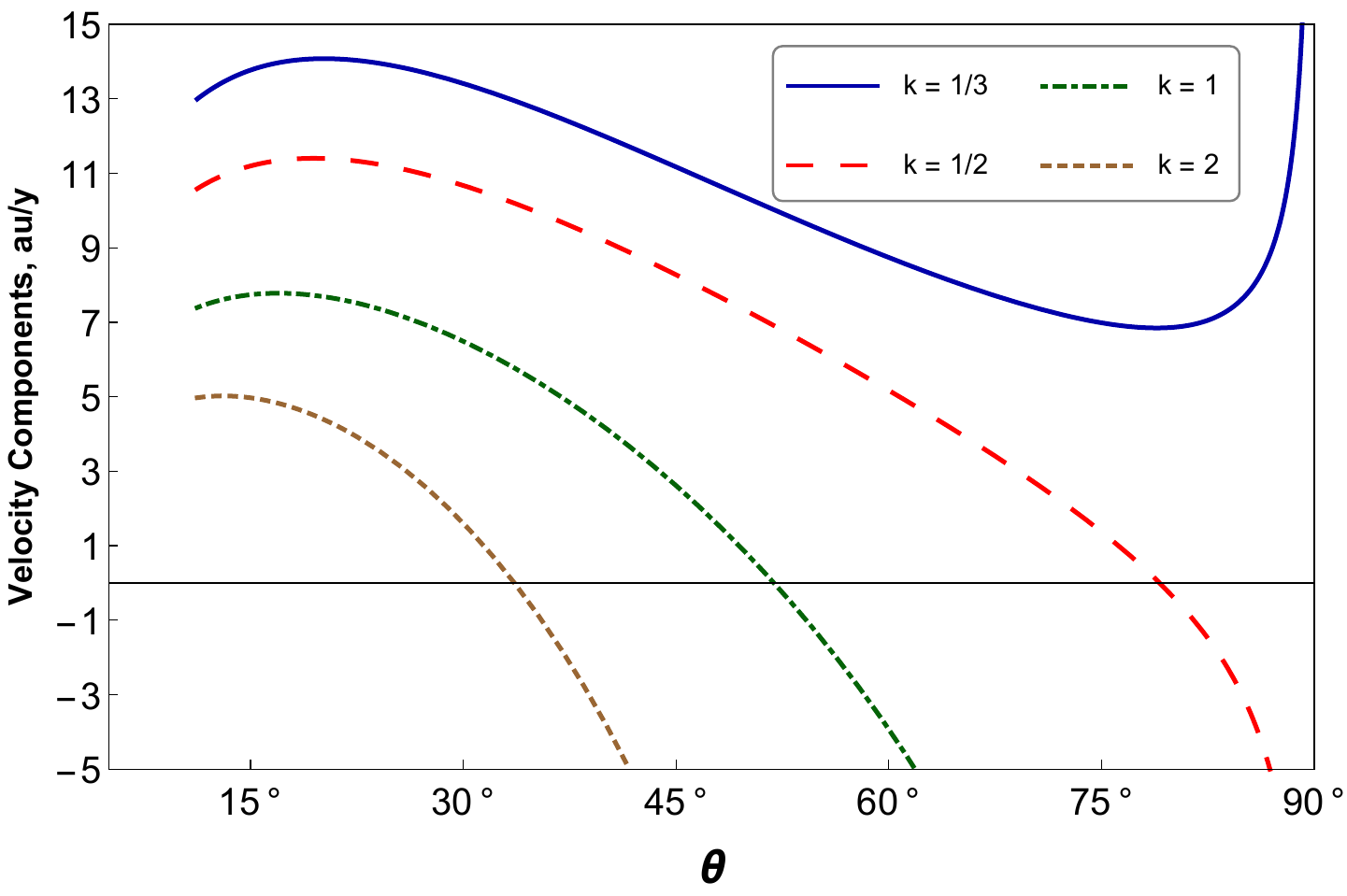}
		\caption{Azimuthal Component}
		\label{fig:vphi_k}
	\end{subfigure}
	\caption{The radial, polar, and azimuthal components of the velocity for a solar sailing spacecraft that is constrained to move in a cylinder. Here, we assume that $\alpha=-45^{\circ}$, the sail is initially at $\theta_{0}=10^{\circ}$, and the clock angle varies linearly as $\theta$.}
	\label{fig:cyl_orbits_velocities_periodic_angle}
\end{figure*}

By changing the form of $n_{\theta}$ and $n_{\varphi}$, it is possible to determine a different  family of orbits for the same constraint equation. For example, let us assume that the polar and azimuthal components of the unit normal vector are periodic relative to the polar angle, \textit{i.e.} for some constants $c$ and $k$, 
\begin{equation}
	\label{eq:periodic_clock}
	n_{\theta}=c\cos k\theta; \quad \quad n_{\varphi}=c\sin k\theta
\end{equation}
where $c=\sqrt{1-n_{r}^{2}}>0$. In this case the period for the change in clock angle is given by $2\pi/k$. Substituting, we see that the azimuthal orbit equation becomes
\begin{equation}
	\label{eq:cyl_azm_orb_eqn_periodic}
	\varphi(\theta)=\varphi(\theta_{0})+\int_{\theta_{0}}^{\theta}\d\nu\frac{2\cos k\nu-\int_{\theta_{0}}^{\nu}\d\eta \cos k\eta \cot\eta}{\sin\nu [2\sin k\nu-\int_{\theta_{0}}^{\nu}\d\eta\sin k\nu \cot\nu\cos(\nu-\eta)]}
\end{equation}
and the components of the velocity are the following:
\begin{equation}
	\label{eq:cyl_thetadot_periodic}
	\thetadot=\left[\frac{-\beta\mu n_{r}^{2}}{r^{3}\cot\theta}s\right]^{1/2},
\end{equation}
\begin{equation}
	\label{eq:cyl_rdot_periodic}
	\rdot=\left[\frac{-\beta\mu}{r}n_{r}^{2}s\cot\theta \right]^{1/2}, 
\end{equation}
and
\begin{equation}
	\label{eq:cyl_phidot_periodic}
	\begin{split}
		\phidot=\left[\frac{-\beta\mu n_{r}^{2}}{r^{3}s\cot\theta}\right]^{1/2}\frac{2c\sin k\theta-\int_{\theta_{0}}^{\theta}\d\eta \ c\sin k\eta \cot\eta }{\sin\theta }
	\end{split}
\end{equation}
where
\begin{equation}
	s=2c\cos k\theta-\int_{\theta_{0}}^{\theta}\d\eta\ c\cos k\eta\cot\eta\cos(\theta-\eta).
\end{equation}
The trajectories and the resulting phase space plots as a function of $\theta$ are presented in Figures \ref{fig:cyl_orbits_periodic_clock} and \ref{fig:cyl_orbits_velocities_periodic_angle}. As $k$ decreases, the trajectory becomes more helical since $n_{\theta}$ is much smaller for smaller $k$. Up to first order approximation, $n_{\theta}$ is linear with respect to $\theta$, with slope given by $k$. On the other hand, $n_{\varphi}$ is a constant near 1.  In return, the ratio $n_{\varphi}/n_{\theta}$, which qualitatively describes the magnitude of $\varphi(\theta)$, becomes larger for smaller $k$. Thus, $\varphi(\theta)$ increases for smaller $k$, effectively increasing the number of revolutions of the spacecraft.

A different behavior is observed for $k\ge 1$. In this range, the trajectory of the sail reverses in the $\phihat$ direction after reaching some $\theta$. This is also shown in the azimuthal velocity changing sign as $\theta$ increases. In contrast to the $k<1$ case, due to the period of $n_{\varphi}$ and $n_{\theta}$ being less than $2\pi$, the value of  $n_{\varphi}$ changes sign even at $\theta<90^{\circ}$. For $k\ge2$, the sail will not reach the ecliptic plane at all \textit{i.e.} that orbit equation diverges for some $\theta<90^{\circ}$. This seemingly unphysical solution might be an artifact of the coordinate system having singularities at the ecliptic plane. This problem can be mitigated by a modification in the control law of the sail when it reaches the singularity.  Nevertheless, this family of solutions can have potential applications for rendezvous missions in which the azimuthal displacement of the sail is negative.

\section{Applications to Displaced non-Keplerian Orbits}
\label{sec:displaced-NKO}

\begin{figure*}[h!]
	\centering
	\includegraphics[scale=0.44]{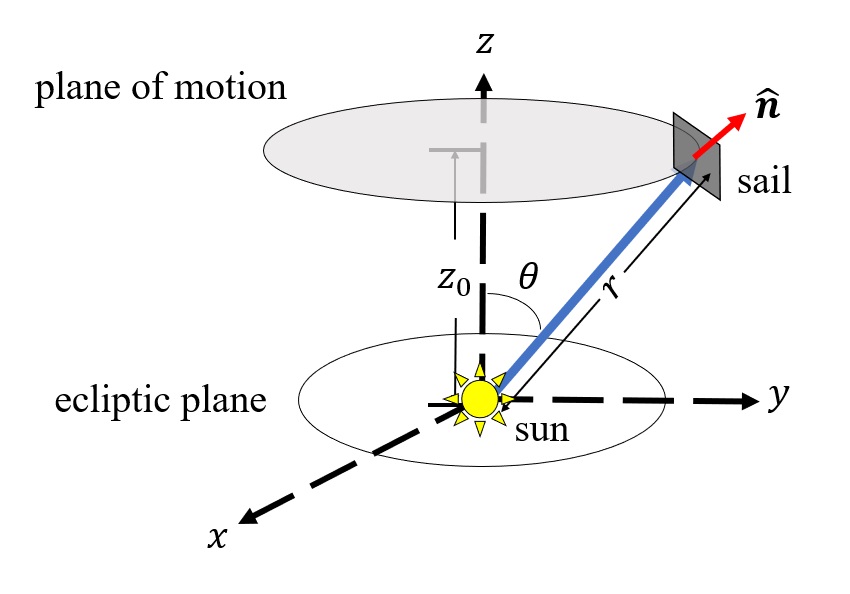}
	\caption{A displaced non-Keplerian orbit.}
	\label{fig:NKO_diagram}
\end{figure*}

We can also recover some displaced, non-Keplerian solutions obtained by McInnes \textit{et al.} \citep{mcinnes1998dynamics,mckay2011survey,mcinnes2004solar} using our proposed method. This can be done by assuming that the sail is constrained on a plane at a distance $z_{0}=r\cos\theta$ from the ecliptic plane (\textit{See Figure }\ref{fig:NKO_diagram}). 
Taking the time derivative and re-writing the resulting expression in the form given by equation \eqref{eq:constraint}, the constraint equation for displaced non-Keplerian orbits is given by
\begin{equation}
	\label{eq:nko_constraint}
	G(\theta)=\tan\theta.
\end{equation}
The radial orbit equation, from the geometry of the problem, is given by
\begin{equation}
	r(\theta)=r_{0}\frac{\sec\theta}{\sec\theta_{0}}
\end{equation}
Similar to what was done in the cylindrical case, one can use equation \eqref{eq:azimuthal_orbit_eqn} to obtain the azimuthal orbit equation and the velocities for a given form of $n_{\theta}(\theta)$ and $n_{\varphi}(\theta)$. 

\subsection{Constant Cone and Clock Angles}

As a special case, closed-form expressions of the azimuthal orbit equation and the components of the velocity can be determined by assuming that both the cone and clock angles are constant throughout the mission. Letting 
\begin{equation}
	C=1+\cos(\nu-\theta_{0})+\ln\left(\frac{1+\tan(\nu/2)}{1-\tan(\nu/2)}\frac{1-\tan(\theta_{0}/2)}{1+\tan(\theta_{0}/2)}\right),
\end{equation}
we see that the azimuthal orbit equation for the solar sail in a non-Keplerian orbit with a constant cone and clock angle is given by
\begin{equation}
	\varphi(\theta)=\varphi(\theta_{0})+\tan\delta\int_{\theta_{0}}^{\theta}\d\nu\frac{\left(2+\ln\frac{\sec\nu}{\sec\theta_{0}}\right)\csc\nu}{C}
\end{equation}
Substituting equation \eqref{eq:nko_constraint} to \eqref{eq:thetadot}, \eqref{eq:rdot} and \eqref{eq:phidot}, we see that
\begin{equation}
	\label{eq:nko_thetadot}
	\thetadot=\left[\frac{-\tbeta n_{r}^{2}n_{\theta}}{r^{3}\tan\theta}C\right]^{1/2},
\end{equation}
\begin{equation}
	\label{eq:nko_rdot}
	\rdot=\left(-\frac{\tbeta}{r}n_{r}^{2}n_{\theta}\ C\tan\theta\right)^{1/2},
\end{equation}
and
\begin{equation}
	\label{eqn:nko_phidot}
	\phidot=\left[\frac{-\tbeta }{r^{3}\tan\theta }\frac{n_{r}^{2}n_{\varphi}^{2}}{n_{\theta}}\right]^{1/2}\frac{2+\ln\frac{\sec\nu}{\sec\theta_{0}}}{C^{1/2}\sin\theta}.
\end{equation}
The resulting plots of the trajectory and the phase space plots are shown in Figure \ref{fig:nk_orbits_constant_cone_angle}:

\begin{figure*}[h!]
	\centering
	\begin{subfigure}{.5\textwidth}
		\centering
		\includegraphics[scale=0.5]{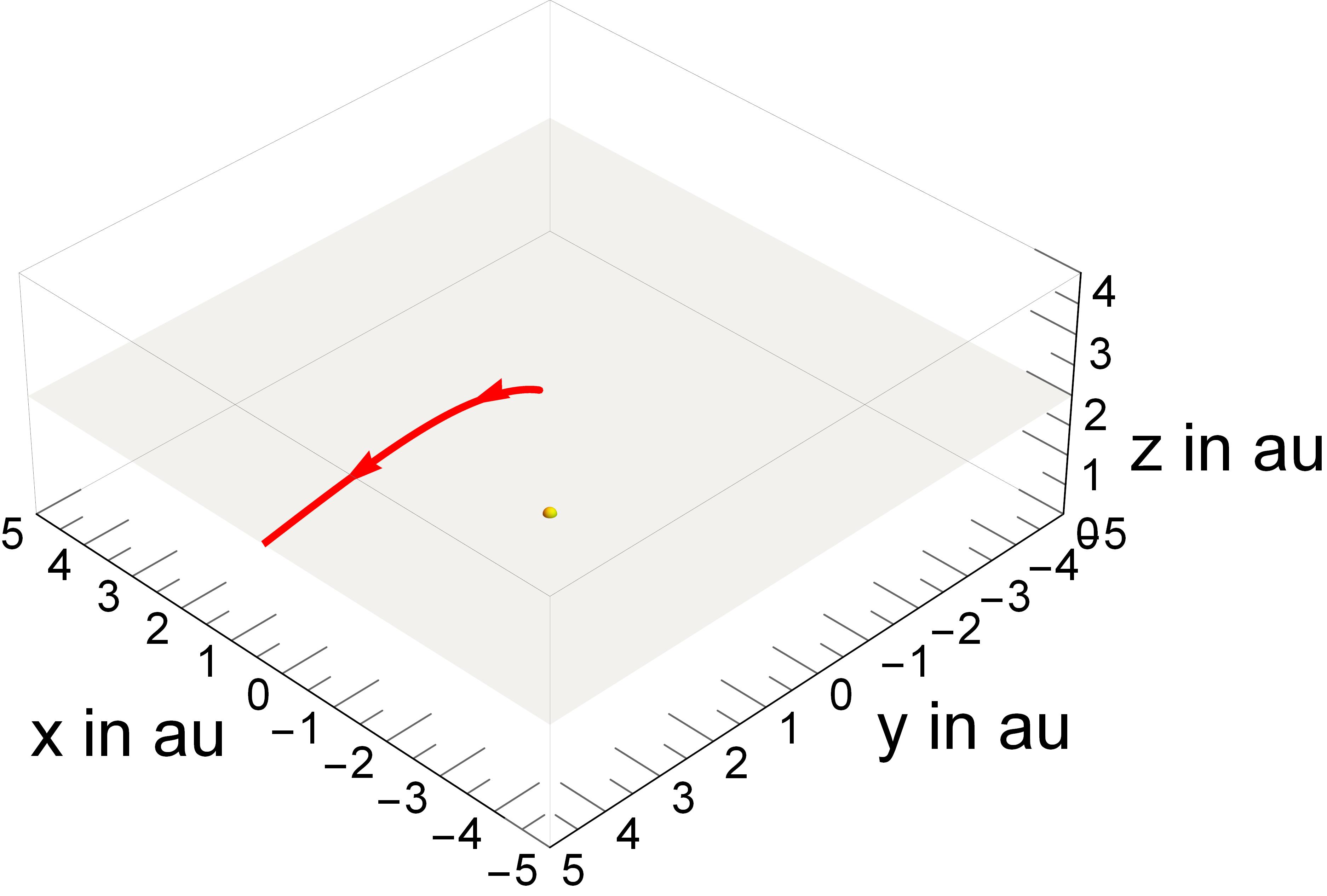}
		\caption{$\delta=30^{\circ}$}
		\label{fig:nko_delta=30}
	\end{subfigure}%
	\begin{subfigure}{.5\textwidth}
		\centering
		\includegraphics[scale=0.5]{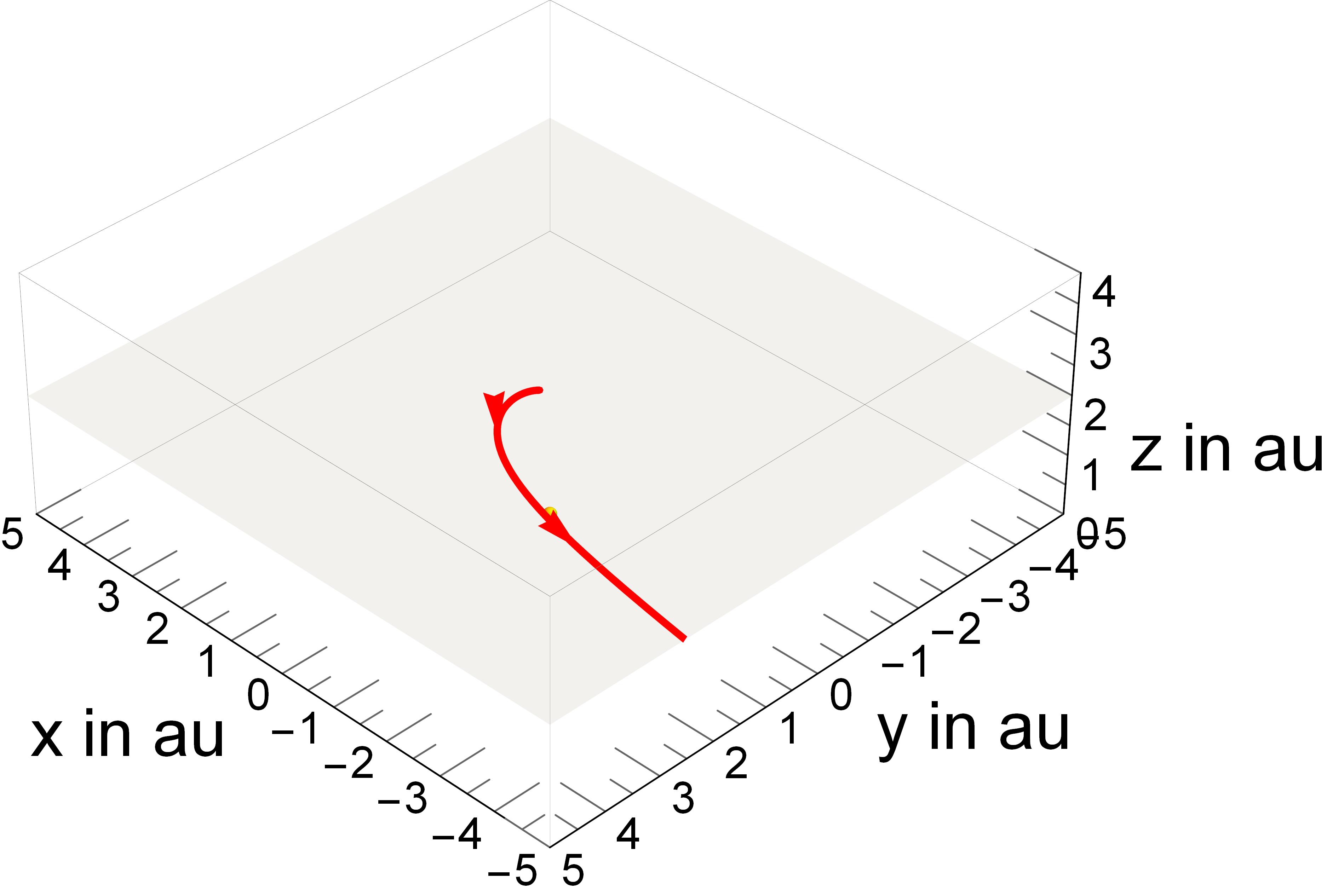}
		\caption{$\delta=45^{\circ}$}
		\label{fig:nko_delta=45}
	\end{subfigure}
	\begin{subfigure}{.5\textwidth}
		\centering
		\includegraphics[scale=0.5]{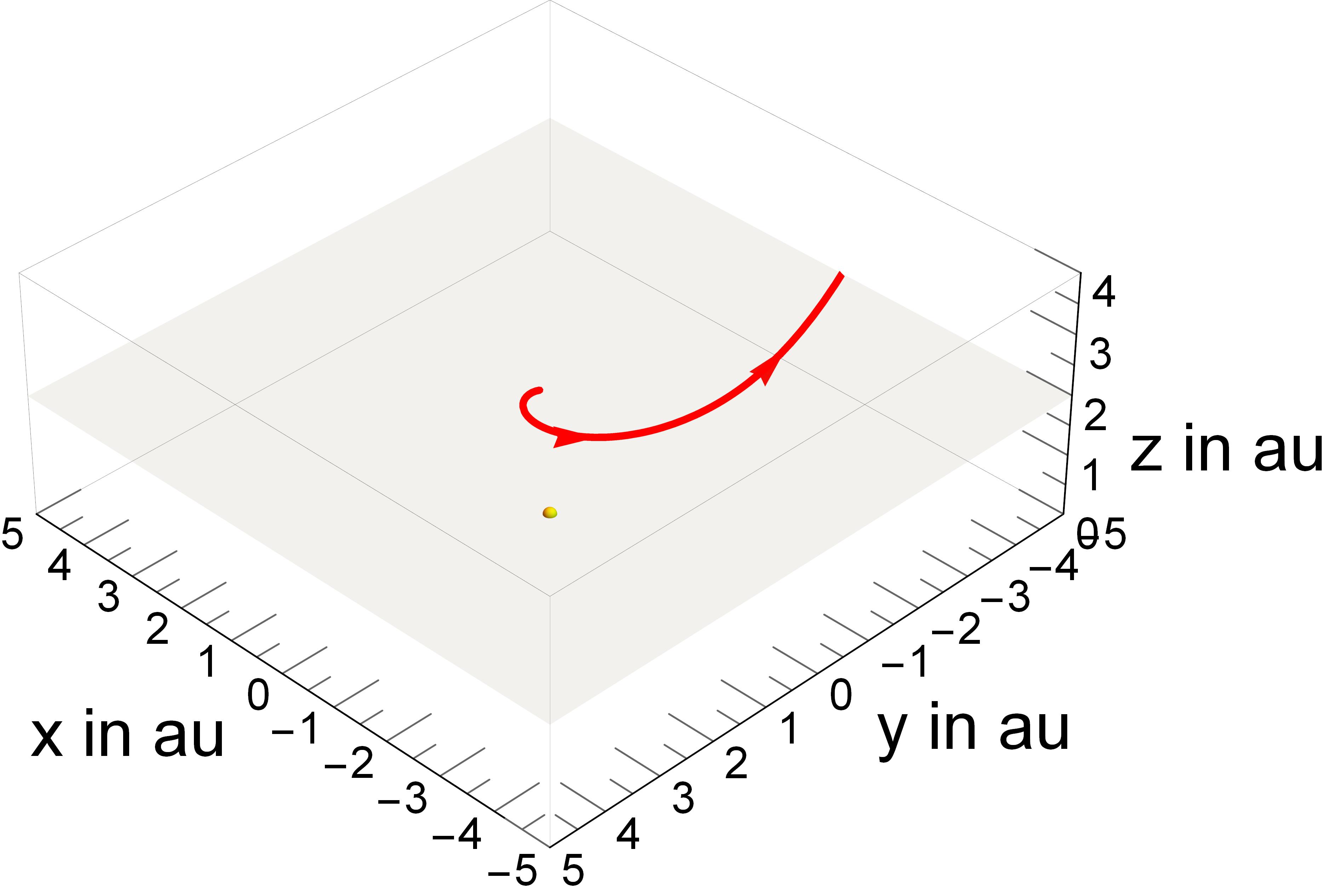}
		\caption{$\delta=60^{\circ}$}
		\label{fig:nko_delta=60}
	\end{subfigure}%
	\begin{subfigure}{.5\textwidth}
		\centering
		\includegraphics[scale=0.5]{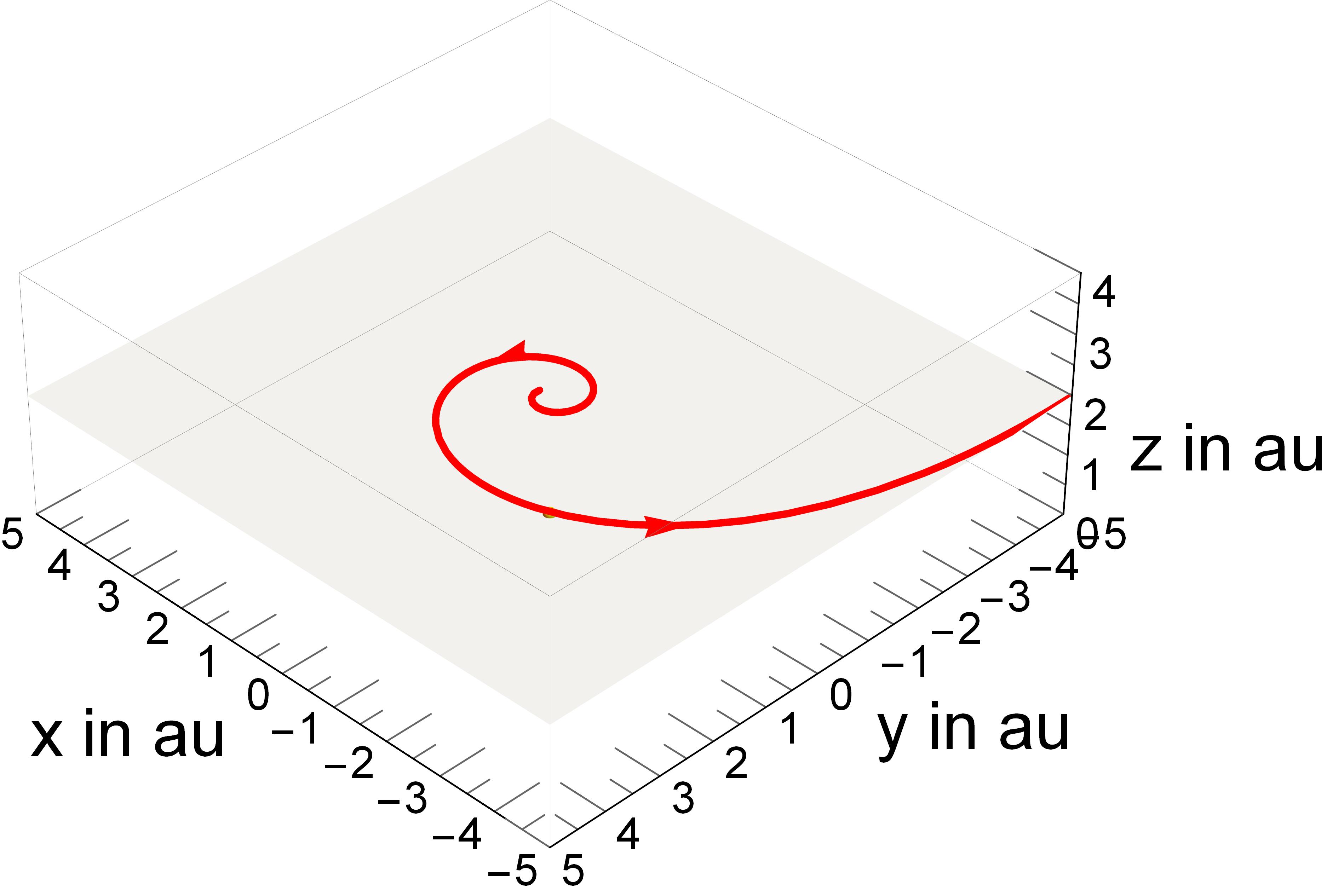}
		\caption{$\delta=75^{\circ}$}
		\label{fig:nko_delta=75}
	\end{subfigure}
	\caption{Displaced non-Keplerian orbits with cone and clock angles constant throughout the mission. We fix the cone angle to be $45^{\circ}$ in the mission.}
	\label{fig:nk_orbits_constant_cone_angle}
\end{figure*}

The trajectories obtained are spirals that are vertically displaced from the ecliptic plane. The shape of these trajectories can be understood by considering the radial, polar and tangential forces acting on the spacecraft. For a receding spacecraft in a displaced non-Keplerian orbit, the radial component of the solar radiation pressure is stronger than the radially-inward gravitational force from the sun, which is opposite with what happens in orbits constrained on cylinders. This results in the radial component of the force being outward. Combining this effect with the polar component of the force gives us a spacecraft constrained on a displaced plane. The azimuthal component of the force causes the solar sail to revolve about the axis perpendicular to the ecliptic, which increases as the clock angle increases. 

For the sail to approach the sun in a displaced non-Keplerian orbit, the force due to solar radiation pressure must be inverted, which can be achieved by reversing the direction of both the cone and the clock angles. In return, the radial component of the solar radiation pressure is now radially-inward and hence parallel to the direction of the gravitational force. 

The nature of the orbits being receding with a constant vertical displacement from the sun makes it viable for missions towards bodies with highly-eccentric orbits and those bodies outside the asteroid belt.

\subsection{Periodic Cone and Clock Angles}
\begin{figure*}[h!]
	\centering
	\begin{subfigure}{0.5\textwidth}
		\centering
		\includegraphics[scale=0.5]{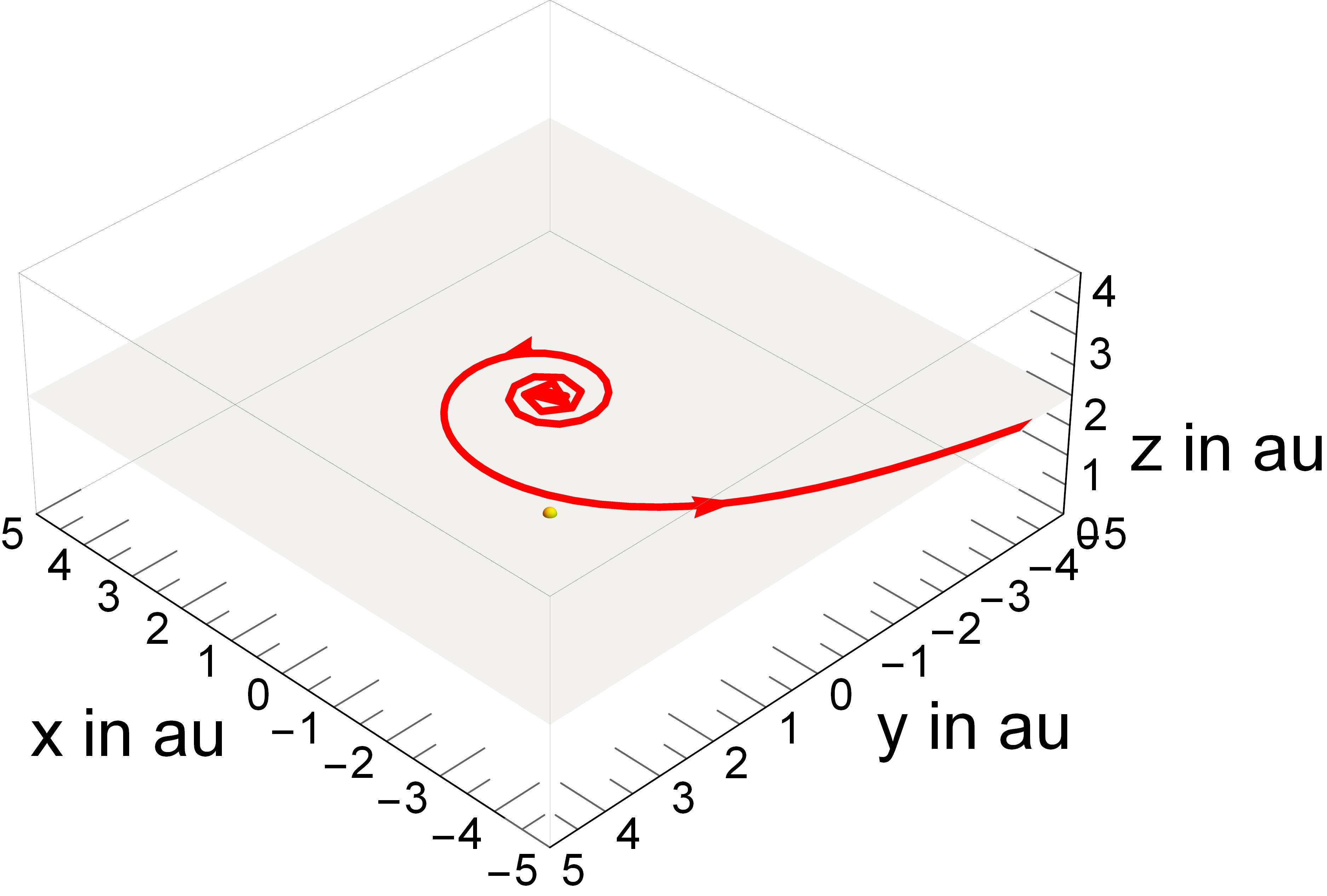}
		\caption{$k=1/3$}
		\label{fig:nko_1/3}
	\end{subfigure}%
	\begin{subfigure}{0.5\textwidth}
		\centering
		\includegraphics[scale=0.5]{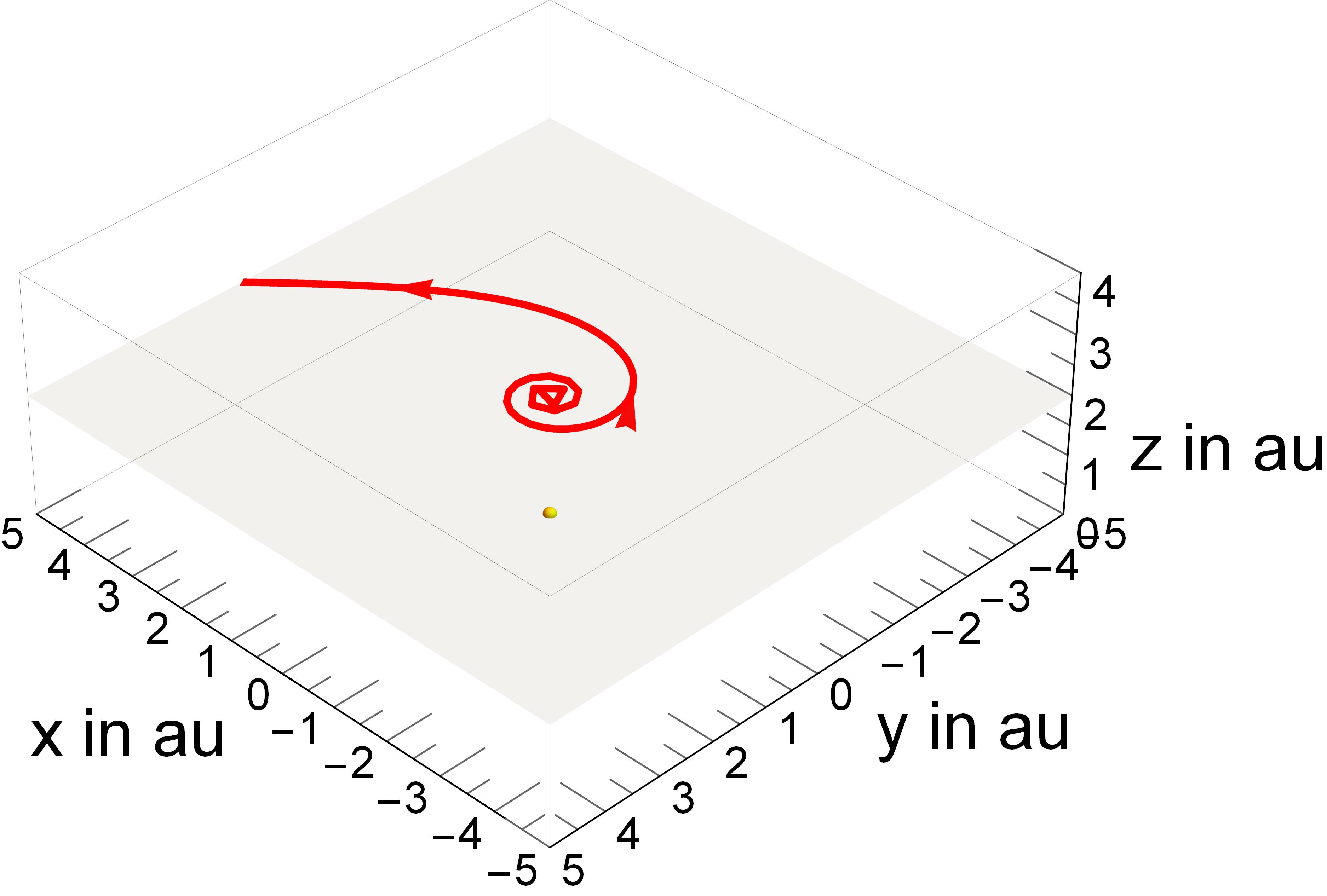}
		\caption{$k=1/2$}
		\label{fig:nko_k=1-2}
	\end{subfigure}
	\begin{subfigure}{0.5\textwidth}
		\centering
		\includegraphics[scale=0.5]{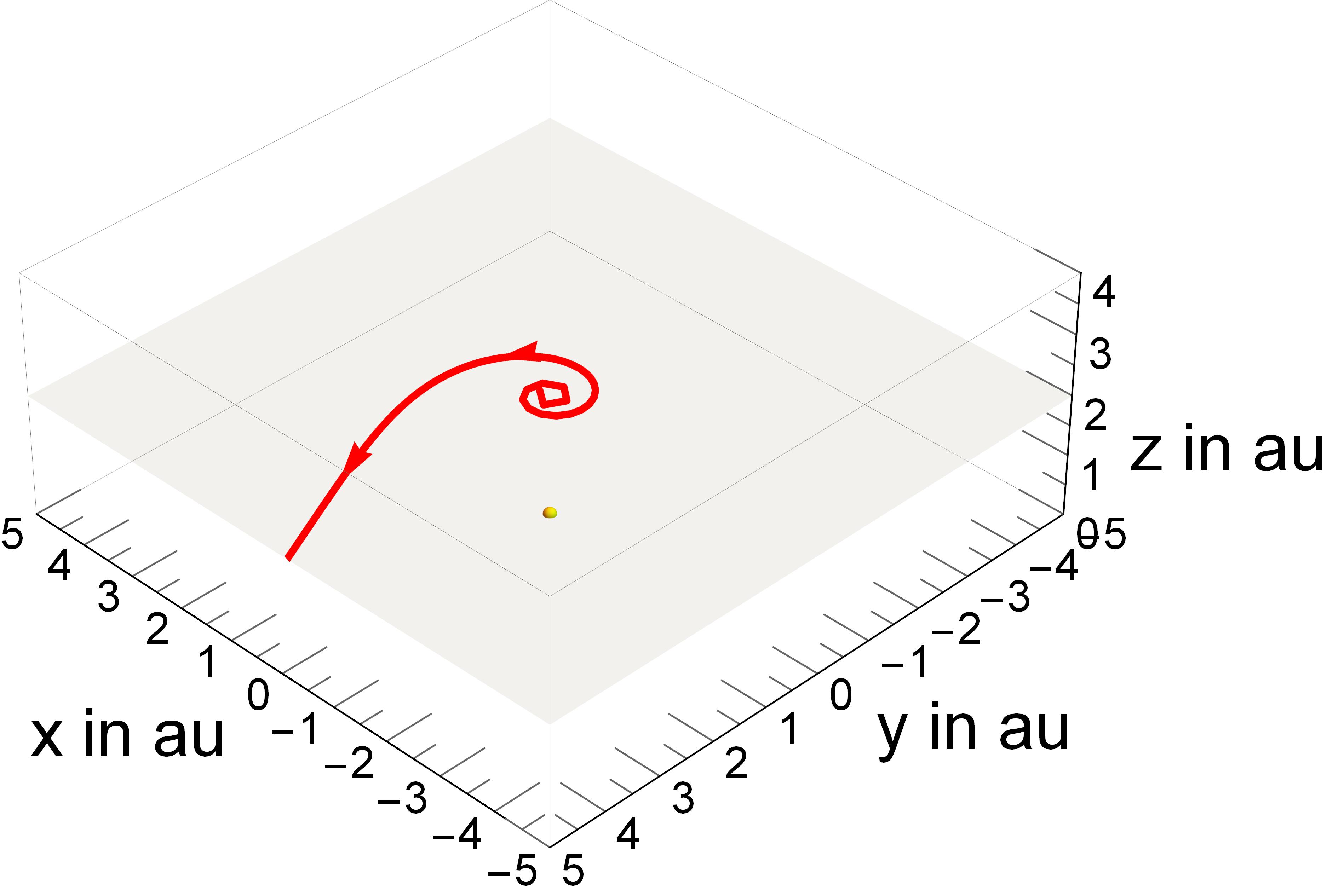}
		\caption{$k=2/3$}
		\label{fig:nko_k=2-3}
	\end{subfigure}%
	\begin{subfigure}{0.5\textwidth}
		\centering
		\includegraphics[scale=0.5]{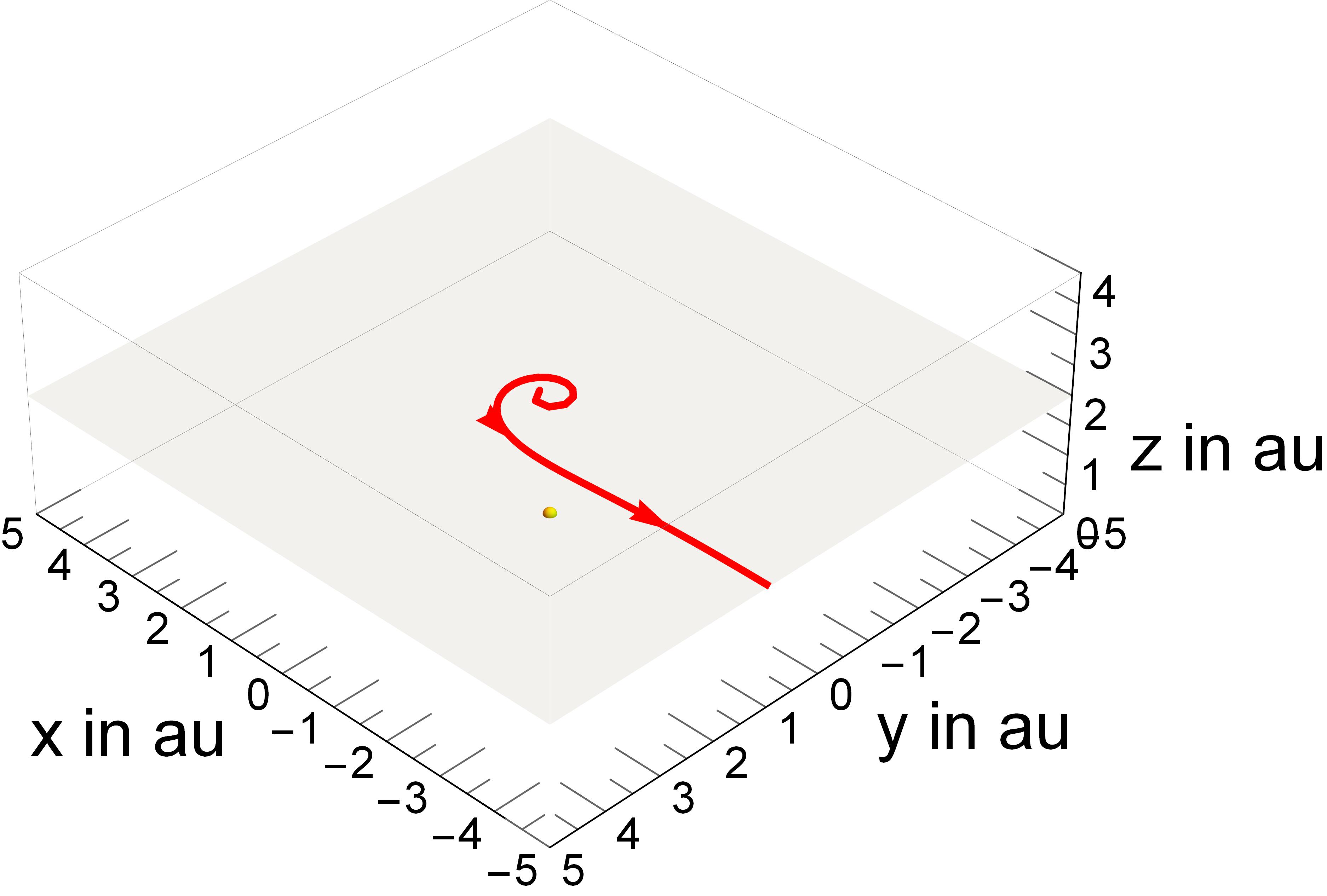}
		\caption{$k=1$}
		\label{fig:nko_k=1-0}
	\end{subfigure}
	\begin{subfigure}{0.5\textwidth}
		\centering
		\includegraphics[scale=0.5]{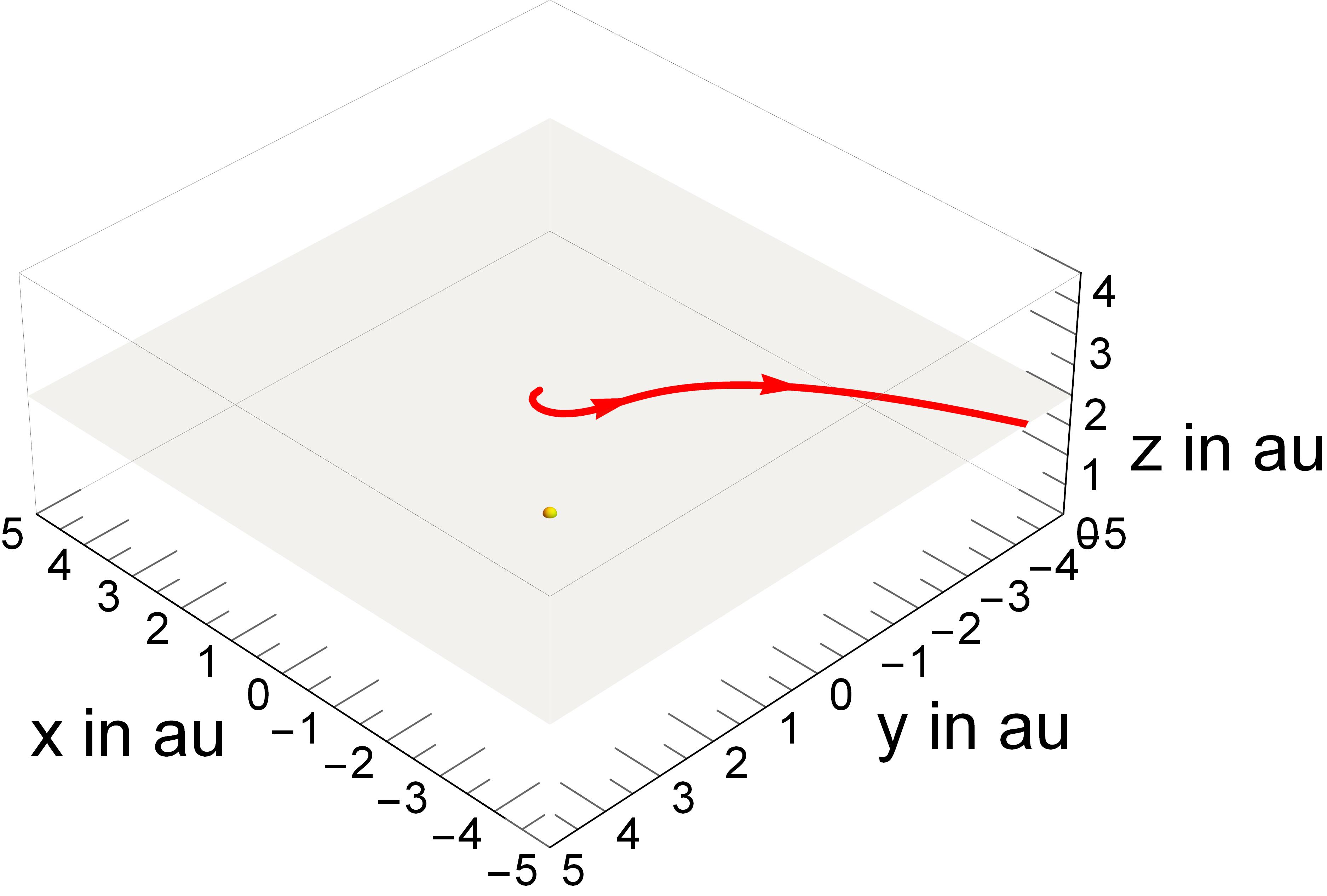}
		\caption{$k=2$}
		\label{fig:nko_k=2-0}
	\end{subfigure}%
	\begin{subfigure}{0.5\textwidth}
		\centering
		\includegraphics[scale=0.5]{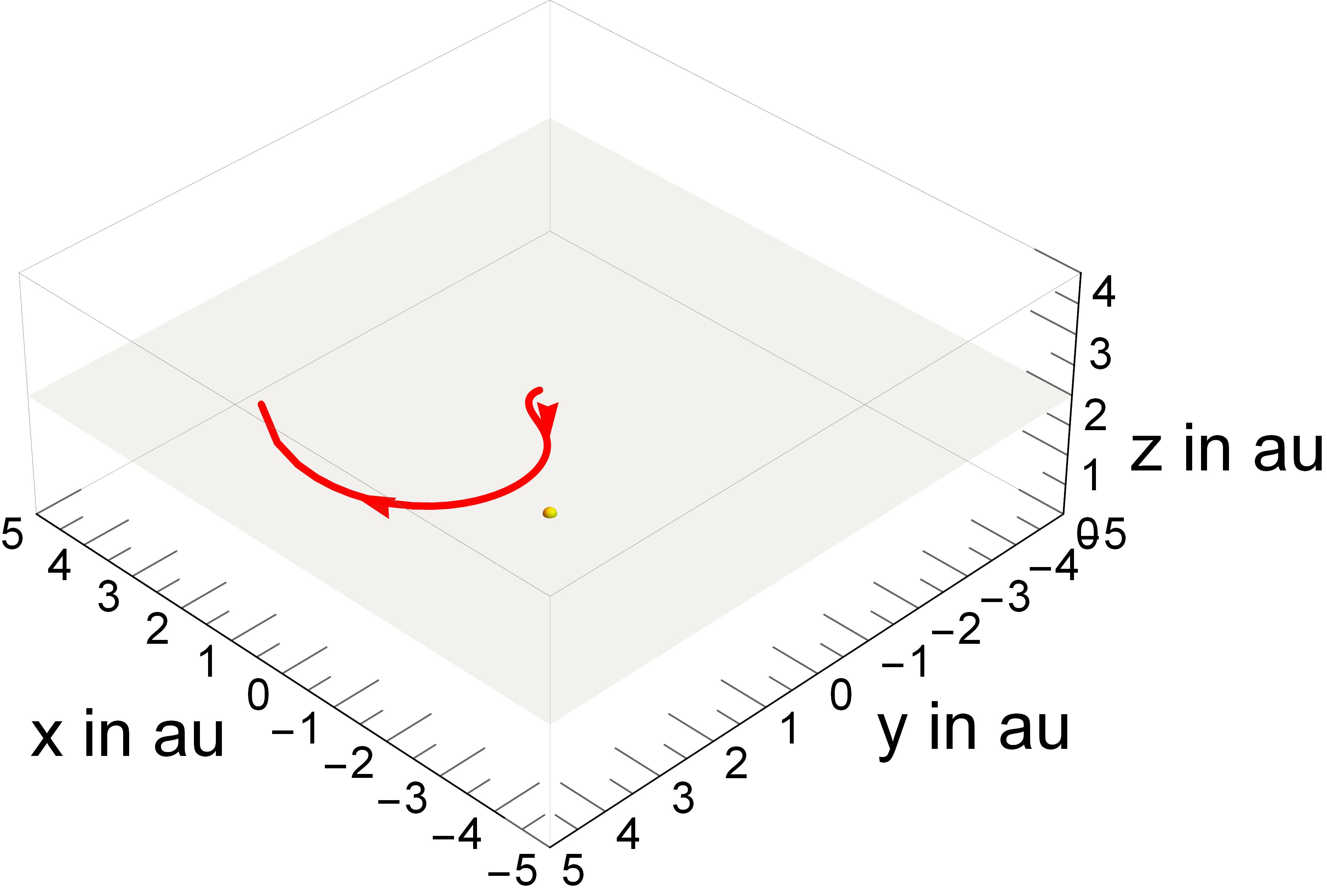}
		\caption{$k=3$}
		\label{fig:nko_k=3-0}
	\end{subfigure}
	\caption{Displaced non-Keplerian orbits with periodic clock angles. The initial cone angle is set at $\alpha=45^{\circ}$. The sun is specified to be at the origin of the coordinate system.}
	\label{fig:nko_orbits_periodic_clock}
\end{figure*}

We can also consider displaced non-Keplerian orbits with periodic polar and azimuthal components of the unit vector \textit{i.e.} $n_{\theta}$ and $n_{\varphi}$ following equation \eqref{eq:periodic_clock}. In this case, the azimuthal orbit equation becomes
\begin{equation}
	\label{eq:nko_azm_orb_eqn_periodic}
	\varphi(\theta)=\varphi(\theta_{0})+\int_{\theta_{0}}^{\theta}\d\nu\frac{2\cos k\nu+\int_{\theta_{0}}^{\nu}\d\eta \cos k\eta \tan\eta}{\sin\nu [2\sin k\nu+\int_{\theta_{0}}^{\nu}\d\eta\sin k\nu \tan\nu\cos(\nu-\eta)]}.
\end{equation}
The trajectories for different values of $k$ are shown in Figure \ref{fig:nko_orbits_periodic_clock}. Similar to what was observed in orbits constrained in cylinders, for $k<1$ the number of revolutions increases as $k$ decreases. However, there is an upper bound for $\theta$ in which the spiraling stops and the trajectory follows a straight line that goes up to infinity. 

On the other hand, while the displaced non-Keplerian orbits obtained are still receding from the sun, the direction of the $\phihat$ reverses for $k>1$. The reversal of the azimuthal coordinate becomes more apparent as $k$ increases such that at $k=3$, only a small part of the trajectory has positive $\varphi$. Similar to the case in orbits constrained on cylinders, the change in the direction of the azimuthal position is related to the nature of $n_{\theta}$ changing signs for $\theta<90^{\circ}$. 

At $k>1$, the terminal position of the polar angle becomes lesser than $90^{\circ}$ as $k$ increases. In fact, $\theta_{f}=60^{\circ}$ for $k=3$. Since in the case of periodic cone and clock angles, $\delta=k\theta$, the period for $n_{\theta}$ and $n_{\varphi}$ is less than $360^{\circ}$ for $k>1$. Consequently, it will also take $\theta<90^{\circ}$ for $k\theta$ to reach $180^{\circ}$, which is the upper bound of the range of values for the clock angle $\delta$.

\section{Conclusions}

In this paper, we have provided an alternative approach of designing the trajectories of a solar sailing spacecraft. With the use of a generalized Laplace-Runge-Lenz vector and through simple assumptions in the orbit's geometry, we have solved the equation of motion of a solar sail by obtaining its radial and azimuthal orbit equations. We have shown that for the case when the radial component of the velocity is related to its polar component by a surface constraint, a conserved quantity can be derived, provided the cone angle is constant throughout the mission. The first integral of motion is used to check what values of the lightness number, the clock angles, and other controllable parameters can define a physically realizable orbit. From the orbit equations, we also determined an analytic expression for the sail's velocity, which is vital in analyzing the solar sail's motion.

The method is used to determine the trajectories of orbits with cylindrical and displaced planar surface constraints. We have shown that if the clock angle increases linearly with the polar angle, hence giving us a periodic $n_{\theta}$ and $n_{\varphi}$, a sail traversing a cylindrical surface may rendezvous along the azimuthal direction, depending on the proportionality constant. In fact, any combination of $n_{\theta}$ and $n_{\varphi}$ can be used as a control law as long as the normalization condition is satisfied. Other ways of changing the clock angle have not yet been studied and can be a subject of future research. For example, we assumed that the time for the sail to change its clock angle is almost instantaneously.  In practice however, the switching time and frequency matter in optimization problems that analyzing their effect can be a topic for future research \citep{ceriotti2021simple}.

Our surface constraint approach adds to the various semi-analytic methods that can be used in designing the trajectories of a solar sailing spacecraft. Because of its simplicity, we see the method being vital in solar sailing trajectory optimization. The surface constraint approach we proposed can give us a family of initial guesses in the orbit equation solution, which can then be used in numerical trajectory optimization. We see that our method is applicable in missions involving asteroids and exotic heavenly bodies for which a solar sail can be deployed.

\bibliographystyle{jasr-model5-names}
\biboptions{authoryear}
\bibliography{garrido-esguerra-manuscript-source-file}

\end{document}